%% file: main.tex
\documentclass[onefignum,onetabnum]{siamart190516}

\usepackage{mathtools}
\usepackage{stmaryrd}

\usepackage{yhmath}
\usepackage{url}
\usepackage{latexsym,epsfig,amssymb,graphics,tabularx}
\usepackage{color,graphicx}
\usepackage{amsfonts}
\usepackage{graphpap}
\usepackage{listings}
\usepackage{booktabs}
\usepackage{multirow}
\usepackage{enumitem}
\usepackage{fancyvrb}

\usepackage{algpseudocode}
\algrenewcommand\algorithmicrequire{\textbf{Input:}}
\algrenewcommand\algorithmicensure{\textbf{Output:}}
\algnewcommand{\LineComment}[1]{\State \(\triangleright\) #1}

\if@review
\usepackage{etoolbox} 
\makeatletter 
\newcommand*\linenomathpatch{\@ifstar{\linenomathpatch@AMS}{\linenomathpatch@}}
\newcommand*\linenomathpatch@[1]{
	\expandafter\pretocmd\csname #1\endcsname {\linenomathWithnumbers}{}{}
	\expandafter\pretocmd\csname #1*\endcsname{\linenomathWithnumbers}{}{}
	\expandafter\apptocmd\csname end#1\endcsname {\endlinenomath}{}{}
	\expandafter\apptocmd\csname end#1*\endcsname{\endlinenomath}{}{}
}
\newcommand*\linenomathpatch@AMS[1]{
	\expandafter\pretocmd\csname #1\endcsname {\linenomathWithnumbersAMS}{}{}
	\expandafter\pretocmd\csname #1*\endcsname{\linenomathWithnumbersAMS}{}{}
	\expandafter\apptocmd\csname end#1\endcsname {\endlinenomath}{}{}
	\expandafter\apptocmd\csname end#1*\endcsname{\endlinenomath}{}{}
}
\let\linenomathWithnumbersAMS\linenomathWithnumbers
\patchcmd\linenomathWithnumbersAMS{\advance\postdisplaypenalty\linenopenalty}{}{}{}
\makeatother 

\linenomathpatch{equation}
\linenomathpatch*{gather}
\linenomathpatch*{multline}
\linenomathpatch*{align}
\linenomathpatch*{alignat}
\linenomathpatch*{flalign}
\fi

\usepackage{pgf,tikz}
\usepackage{epstopdf}
\usetikzlibrary{shapes,arrows,automata}

\ifpdf
\DeclareGraphicsExtensions{.eps,.pdf,.png,.jpg}
\else
\DeclareGraphicsExtensions{.eps}
\fi

\input{def}

\headers{On Termination of MPPs with Equality Conditions}{Y. Li, N. Zhan, M. Chen, H. Lu, G. Wu, and J.-P. Katoen}

\title{On Termination of Polynomial Programs with Equality Conditions
}

\author{Yangjia Li\footnotemark[2]\ \footnotemark[3]
	\and Naijun Zhan\footnotemark[2]
	\and Mingshuai Chen\footnotemark[4]
	\and Hui Lu\footnotemark[5]
	\and Guohua Wu\footnotemark[6]
	\and Joost-Pieter Katoen\footnotemark[4]}

\usepackage{amsopn}

\ifpdf
\hypersetup{
  pdftitle={On Termination of Polynomial Programs with Equality Conditions},
  pdfauthor={Y. Li, N. Zhan, M. Chen, H. Lu, G. Wu, and J.-P. Katoen}
}
\fi

\makeatletter
\newcommand{\leqnomode}{\tagsleft@true}
\newcommand{\reqnomode}{\tagsleft@false}
\makeatother

\begin{document}

\maketitle

\renewcommand{\thefootnote}{\fnsymbol{footnote}}

\footnotetext[2]{State Key Laboratory of Computer Science, Institute of Software, Chinese Academy of Sciences \& University of Chinese Academy of Sciences, Beijing, China 
	(\email{znj@ios.ac.cn}).}
\footnotetext[3]{University of Tartu, Tartu, Estonia 
	(\email{yangjia.li@ut.ee}).}
\footnotetext[4]{
	RWTH Aachen University, Aachen, Germany 
	(\{\href{mailto:chenms@cs.rwth-aachen.de}{chenms},\href{mailto:katoen@cs.rwth-aachen.de}{katoen}\}\textcolor{siamexlinkcolor}{@cs.rwth-aachen.de}).}
\footnotetext[5]{Nanjing Audit University, Nanjing, China 
	(\email{luhui@nau.edu.cn}).}
\footnotetext[6]{Nanyang Technological University, Singapore 
	(\email{guohua@ntu.edu.sg}).}

\renewcommand{\thefootnote}{\arabic{footnote}}

\input{abstract}

\input{introduction}

\input{overview}

\input{preliminaries}

\input{bound}

\input{mpps}

\input{decidability}

\input{extensions}

\input{experiments}

\input{conclusion}



\bibliographystyle{siamplain}
\bibliography{references}
\end{document}

%% file: abstract.tex
\begin{abstract}
	We investigate the termination problem of a family of multi-path polynomial programs (MPPs), in which all assignments to program variables are polynomials, and test conditions of loops and conditional statements are polynomial equalities. We show that the set of non-terminating inputs (NTI) of such a program is algorithmically computable, thus leading to the decidability of its termination. To the best of our knowledge, the considered family of MPPs is hitherto the largest one for which termination is decidable. We present an explicit recursive function which is essentially Ackermannian, to compute the maximal length of ascending chains of polynomial ideals under a control function, and thereby obtain a complete answer to the questions raised by Seidenberg~\cite{Seidenberg1971}. This maximal length facilitates a precise complexity analysis of our algorithms for computing the NTI and deciding termination of MPPs. We extend our method to programs with polynomial guarded commands and show how an incomplete procedure for MPPs with inequality guards can be obtained. An application of our techniques to invariant generation of polynomial programs is further presented.
\end{abstract}



%% file: introduction.tex
\section{Introduction}
Termination analysis \cite{cook:termination,yang:advances} has been a long-standing, active field of research due to its great importance in program verification. However, the termination problem of programs is equivalent to the well-known halting problem \cite{turing:computable}, and hence is undecidable in general. A complete method for termination analysis for programs, even for linear or polynomial programs, is therefore impossible \cite{tiwari:terminate,Muller-Olm,bradley:polynomial,braverman:terminate}.
Consequently, a practical way is to provide sufficient conditions for termination and/or nontermination. The classical method for establishing termination of a program uses the so-called \emph{ranking function} that maps the state space of the program to a well-founded domain, thus providing a sufficient condition for termination, see, e.g., \cite{Ben-AmramG14,colon:synthesis,podelski:ranking,podelski:transition,chen:ranking,cousot:lagrangian}. In \cite{gupta:NT}, a sufficient condition for identifying non-terminating inputs was presented for linear programs; while in \cite{Harris10}, sufficient conditions for finding terminating and non-terminating inputs were investigated, as well as techniques for checking the two conditions in parallel for termination analysis.
\oomit{For more general programs, many other techniques, like predicate abstraction, parametric abstraction, fair assumption, Lagrangian relaxation, semidefinite programming, sum of squares, curve fitting, and so on, have been successfully applied \cite{cousot:termination,cousot:lagrangian}.}

In contrast, Tiwari \cite{tiwari:terminate} noticed that termination of a class of simple linear loops is related to the eigenvalues of the assignment matrix and proved that the termination problem of these linear programs with inputs over the reals $\mathbb{R}$ is decidable. The same line of theory was further developed in \cite{braverman:terminate,DBLP:conf/soda/OuakninePW15,DBLP:journals/corr/abs-1302-2762,DBLP:conf/cav/FrohnG19,DBLP:conf/sycss/RebihaMM13,XYZZ2011,XZ2010}. In particular, Bradley et al.~\cite{bradley:polynomial} studied the termination problem of a family of polynomial programs modelled as \emph{multi-path polynomial programs} (MPPs) by using \emph{finite difference trees} (FDTs). These MPP programs cover an expressive class of loops with multiple paths, polynomial loop guards and assignments, which enables practical code abstraction and analysis. Bradley et al.~(ibid.)~proved that the termination problem of MPPs is undecidable. 
A similar idea was employed in \cite{cook:polynomial} for the termination analysis of polynomial programs. Moreover, \cite{LXZZ14} provides sufficient conditions for establishing (non)termination of a subclass of MPPs with equational loop guards. This is achieved by under/over-approximating the set of non-terminating inputs. As in \cite{Harris10}, termination analysis can be conducted by checking these conditions in parallel.  
Until now it remains as an open problem whether 
the termination problem of MPPs with equality guards is decidable.

In this article, we answer the question affirmatively: we show that the termination problem of MPPs with equality guards is decidable. To the best of our knowledge, this family of (multi-path) polynomial programs should be the largest one with a decidable termination problem so far: any extension of it by allowing inequalities or inequations ($\neq$) in the guards may render the termination problem undecidable. Note that program termination with inequality conditions is hard to decide even for simple linear loops, since it is equivalent to the Skolem's problem \cite{OW15} that remains open. The key idea of our approach is as follows:  Given an MPP $P$ with $\ell$ paths, for any input $\bx \in \mathbb{R}^d$, if in the first iteration $\bx$ satisfies the loop guard, then one of the paths in the loop body will be nondeterministically selected and the corresponding assignment will be used to update the value of $\bx$. This results in $\ell$ possible values of $\bx$. Afterwards, the above procedure is repeated until no guard holds anymore.
Thus the execution of an MPP on input $\bx \in \mathbb{R}^d$ forms a tree.
An input $\bx$ is \emph{non-terminating} if the execution tree on $\bx$ has an infinite path. From an algebraic perspective, each of such paths forms an ascending chain of polynomial ideals, and an input $\bx$ is non-terminating iff $\bx$ is in the variety of the ascending chain. This can be determined in finitely many steps due to Hilbert's basis theorem \cite{Atiyah1969} guaranteeing that such an ascending chain is always finite. The difficulty is that the execution tree may contain infinitely many paths of distinct lengths, whereas a uniform upper bound of them, say $k$, is required. 
We show how to obtain such a bound. Thus we only need to symbolically execute the loop at most $k$ steps and check whether the given input is a solution to the ideal formed by some path in the induced finite execution tree.

The main technique we propose to decide termination ---including a precise complexity analysis--- of MPPs is to compute an upper bound on the length of the longest path in their execution tree. This is reduced to computing the maximal length of the Hilbert ascending chains under specific conditions. Briefly speaking, a Hilbert ascending chain is of the form $I_0\subset I_1\subset \cdots \subset I_n\subset \cdots$, where each $I_n$ is an ideal of the polynomial ring (over program variables, in our setting). Seidenberg studied in \cite{Seidenberg1971,seidenberg1972constructive} how to obtain an upper bound on these lengths under a \emph{control function} $f$, such that for each $n$, $f(n)$ is an upper bound on the degrees of polynomials in some basis of $I_n$. Seidenberg (ibid.) proved the existence of such a bound in terms of $f$ and the number $d$ of indeterminants of the polynomial ring. Two questions were posed in \cite{Seidenberg1971}:
\begin{itemize}[leftmargin=9.2mm]
	\item[-] Explicitness: \textit{``[...] it would be desirable to bring to a more satisfactory expression [...]''};
	\item[-] Complexity: \textit{``Even for $d\ge3$, where primitive recursiveness looks doubtful [...]''}.
\end{itemize}
The first question asks for a more explicit expression of the bound, since the original one is too complex to compute. The second question is on the complexity of the length, which was conjectured not to be primitive recursive. Our techniques answer both questions in a constructive way.

The main contributions of this work can be summarized as follows.

\begin{itemize}
\item We construct a Hilbert ascending chain of the maximal length, and express its length (i.e., the optimal upper bound) as a recursive function $L(d,f)$ w.r.t.~the control function $f$ and the number $d$ of indeterminants. As a direct consequence of the expression, the maximal length is shown to be Ackermannian, i.e., not primitive recursive. This confirms the complexity conjectured by Seidenberg in \cite{Seidenberg1971}.
\item We show the decidability of the termination problem of multi-path polynomial programs with equational conditions. Algorithms for computing the set of non-terminating inputs (and thus deciding termination) are proposed and demonstrated on a collection of examples from the literature. A precise complexity analysis of the decision algorithms is presented using the obtained maximal length of Hilbert ascending chains.
\item We extend our method to programs with polynomial guarded commands and show how an incomplete procedure for MPPs with inequality guards can be obtained. We further show its application in discovering loop invariants for polynomial programs, which is another fundamental problem in program verification.
\end{itemize}

\subsection{Related work}
In the literature, various well-established techniques on termination analysis are tailored to linear programs, i.e., programs with linear guards and assignments.
For single-path linear programs, Col\'{o}n and Sipma utilized polyhedral cones to synthesize linear ranking functions \cite{colon:synthesis}. Podelski and Rybalchenko~\cite{podelski:ranking}, based on Farkas' lemma, presented a complete method to find linear ranking functions if they exist. Ben{-}Amram and Genaim \cite{Ben-AmramG14} extended the above results in the following way: first, they proved that synthesizing linear ranking functions for single-path linear programs is in co-NP if program variables are interpreted over the integers $\mathbb{Z}$ (in contrast to the PTIME complexity
 over the rationals $\mathbb{Q}$ or reals $\mathbb{R}$); second, they proposed an approach for synthesizing \emph{lexicographical ranking functions} for multi-path linear programs.

In recent years, there has been an increasing interest in the termination problem of nonlinear programs due to their omnipresence in safety-critical embedded systems.
Bradley et al.~\cite{bradley:polynomial} proposed an approach to proving termination of MPPs with polynomial behaviours over $\mathbb{R}$ through FDTs. A similar idea was exploited in \cite{cook:polynomial} for termination analysis of nonlinear command sequences. Typically, with the development of computer-algebra systems, more and more techniques from symbolic computation, for instance, polynomial ideals \cite{DBLP:journals/fac/RebihaMM15,Sontag,emsoft11}, Gr\"{o}bner bases \cite{sank:GB,muller:invariants}, quantifier elimination \cite{kapur:QE} and recurrence relations \cite{rodriguez:simple,kovacs:psolvable}, have been successfully applied to the verification of programs and control systems. Indeed, these techniques can also be applied to polynomial programs to discover termination or nontermination proofs (cf., e.g., \cite{DBLP:conf/ictac/Li16} for a class of single-path polynomial loops). Chen et al.~\cite{chen:ranking} proposed a relatively complete method (w.r.t.~a given template) for generating polynomial ranking functions over $\mathbb{R}$ by a reduction to semi-algebraic system solving. Gupta et al.~\cite{gupta:NT} proposed a practical method to search for counter-examples of termination, by first generating lasso-shaped \cite{cook:code} candidate paths and then checking the feasibility of the ``lassoes" using constraint solving. \oomit{Velroyen and R\"{u}mmer applied invariants to show that terminating states of a program are unreachable from certain initial states, and then identified these ``bad'' initial states by constraint-solving \cite{Velroyen08}.} \oomit{Brockschmidt et al.~detected non-termination and Null Pointer Exceptions for \textsf{Java Bytecode} by constructing and analyzing termination graphs, and implemented a termination prover \textsf{AProVE} \cite{Brockschmidt11}.}

For more general programs, many other techniques, like predicate abstraction, parametric abstraction, fair assumption, Lagrangian relaxation, semidefinite programming, sum-of-squares optimization,  etc., have been successfully applied \cite{cousot:termination,cousot:lagrangian,cook:conditional}.

The following results are closely related to this paper. Tiwari identified in \cite{tiwari:terminate} a class of simple linear loops and proved that its termination problem is decidable over $\mathbb{R}$. Thereafter, the results were extended to affine loops over $\mathbb{Z}$, where the decidability of termination was established for special cases, e.g., linear loops \cite{braverman:terminate}, loops with at most 4 variables \cite{DBLP:conf/soda/OuakninePW15}, and loops with diagonalizable \cite{DBLP:journals/corr/abs-1302-2762,DBLP:conf/soda/OuakninePW15} and triangular \cite{DBLP:conf/cav/FrohnG19} assignment matrices. Xia and Zhang \cite{XZ2010} investigated an extension of Tiwari's simple linear loops by allowing nonlinear loop conditions and proved that the termination problem is still decidable over $\mathbb{R}$, yet becomes undecidable over $\mathbb{Z}$. Recently, Frohn et al.~\cite{DBLP:journals/corr/abs-1910-11588} proved by a reduction to the \emph{existential theory of the reals} that termination of \emph{triangular weakly nonlinear loops} over $\mathbb{R}$ is decidable, and nontermination over $\mathbb{Z}$ and $\mathbb{Q}$ is semi-decidable. It was further shown in \cite{DBLP:conf/lpar/HarkFG20} that the halting problem (w.r.t.~a given input) is decidable for such loops over any ring that lies between $\mathbb{Z}$ and $\mathbb{R_A}$, where $\mathbb{R_A}$ is the set of all real algebraic numbers. In \cite{bradley:polynomial}, Bradley et al.~proved by a reduction from Diophantine equations that the termination problem of MPPs with inequalities as loop conditions is not even semi-decidable. Additionally, M{\"{u}}ller{-}Olm and Seidl \cite{Muller-Olm} proved that the termination problem of affine programs with equalities and inequalities as guards is undecidable. This article extends the picture by showing the decidability of termination for a family of MPPs with equality conditions.
%

Now we place our work in position with existing efforts in addressing Seidenberg's questions concerning the maximal length of Hilbert ascending chains. Most pertinently, Moreno-Soc\'{i}as has proposed in \cite{GMS92} a constructive method for explicitly representing an upper bound on the length of ascending chains of polynomial ideals. Unfortunately, the obtained bound is incorrect, namely, the constructed chain is in fact not the longest. We show that such a spuriously ``maximal'' length may lead to unsoundness issues when being applied in termination analysis of MPPs. In contrast, inspired by Moreno-Soc\'{i}as's results, we present an essentially different approach to constructing a Hilbert ascending chain of the maximal length. A more technical comparison together with an illustrating example can be found in \cref{sec:bound}.

Apart from that, there have been results dedicated merely to the complexity of the length of Hilbert ascending chains (Seidenberg's second question). McAloon \cite{McAloon84} proved that for a linear control function $f$, $L(d,f_t)$ is not primitive recursive on $t$, with $f_t$ defined by $f_t(n)=f(n+t)$. Figueira et al.~established in \cite{FFSS11}, through Dickson's lemma, the relation between the complexity of $L(d,f_t)$ w.r.t.~$t$ and that of $f$ in the fast-growing hierarchy. These results yield a partial answer to Seidenberg's questions (i.e., on the non-primitive recursiveness) yet without addressing the request for an explicit form of $L(d,f)$. More recently, Steila and Yokoyama \cite{DBLP:journals/apal/SteilaY16} presented an alternative proof by using $H$-closures and exploited termination bounds in terms of reverse mathematics. Compared with existing techniques, our method directly computes the maximal length of the chains rather than providing its lower or upper bounds. Such an explicit expression for $L(d,f)$ enables a straightforward way to show the underlying Ackermannian complexity. More importantly, it delivers insight on how many iterations a loop at most will take until termination, and hence is useful in building a decision procedure for termination. The explicit expression for $L(d,f)$ also fulfils Liu et al.'s aspirations for ``a more precise complexity analysis'' of the decision algorithms, as stated in \cite{LXZZ14}.

\vspace*{\baselineskip}
The remainder of the paper is structured as follows. In \cref{sec:nutshell}, we give an overview of our approach through an example. In \cref{sec:pre}, some preliminaries on computational algebraic geometry are reviewed. \Cref{sec:bound} is dedicated to explicitly computing the length of a strictly Hilbert ascending chain of polynomial ideals in the worst case. In \cref{sec:MPP}, we introduce the model of MPPs with equality guards. In \cref{sec:Deci}, we show the decidability of termination for such MPPs by proposing decision algorithms to compute the set of non-terminating inputs. \Cref{sec:ext} extends the method to more general polynomial programs and the related invariant generation problem. \Cref{sec:case} reports some experimental results. \Cref{sec:conc} concludes the paper.

%% file: overview.tex
\section{An overview example} \label{sec:nutshell}
We use the following simple example to give a bird's eye view of our approach.
\begin{exm}[\Verb!overview!]\label[mpp]{exm:running}
\begin{equation*}\begin{split}
&(x,y) \coloneqq (x_0,y_0);\\
&\textsc{ while}\ (x+y=0)\ \textsc{do}\\
&\quad \textsc{if ? then}\ (x,y) \coloneqq (y^2,2x+y);\\
&\quad \textsc{else}\ (x,y) \coloneqq (2x^2+y-1,x+2y+1);\\
&\textsc{ end \ while}
\end{split}\end{equation*}
Here ``\textsc{?}'' indicates that the condition takes a nondeterministic Boolean value, and thus in each iteration 
one of the two assignments is \textit{nondeterministically} selected. 
Our interest is to decide whether or not for any initial value $(x_0,y_0)$ in the set\footnote{The initial set $V$ can be arbitrary, since our approach computes the whole set of non-terminating inputs.} $V=\{(x,y)\mid x^2+y=0\}$, the program terminates.
\end{exm}

For simplicity, the polynomial of the loop guard is denoted as $G(x,y)\triangleq x+y$, and the two polynomial vectors of the assignments as $\bA_1(x,y)=(y^2,2x+y)$ and $\bA_2(x,y)=(2x^2+y-1,x+2y+1)$. Our approach is to compute the set $D$ of all possible initial values of $(x_0,y_0)$ on which the program may not terminate (a.k.a., diverge). Thus, the termination problem of the program on the set $V$ of inputs can be solved by checking whether $V \cap D = \emptyset$. The procedure is outlined as follows.
\begin{enumerate}
\item Let $D_0$ denote the solutions of $G(x,y)=x+y=0$. Thus, $(x_0,y_0)\in D_0$ means that the loop body is executed at least once.
\item Denote by $D_{(1)}$ and $D_{(2)}$ the solution sets of equations:
\begin{equation*}
\begin{split}
&\left\{
   \begin{aligned}
   &G(x,y)=x+y=0\\
   &G(\bA_1(x,y))=y^2+(2x+y)=0
   \end{aligned}\right.\\
{\rm and}&\\
&\left\{
   \begin{aligned}
   &G(x,y)=x+y=0\\
   &G(\bA_2(x,y))=(2x^2+y-1)+(x+2y+1)=0
   \end{aligned}\right.
\end{split}
\end{equation*}
respectively. Similarly,  $(x_0,y_0)\in D_{(i)}$ ($i \in \{1,2\}$) indicates that the loop body may be executed at least twice by correspondingly choosing $\bA_i$ to be the assignment in the first iteration. Hence, $(x_0,y_0)\in D_1\triangleq D_{(1)}\cup D_{(2)}$ allows at least two iterations in the execution. It is easy to calculate that $D_{(1)}=\{(0,0),(-1,1)\}$, $D_{(2)}=\{(0,0),(1,-1)\}$, and thus $D_1=\{(0,0),(1,-1),(-1,1)\}$.
\item Similarly, the solution set $D_{(ij)}$ ($i,j \in \{1,2\}$) of equation
\begin{equation*}
\left\{
   \begin{aligned}
   &G(x,y)=0\\
   &G(\bA_i(x,y))=0\\
   &G(\bA_j(\bA_i(x,y)))=0
   \end{aligned}\right.
\end{equation*}
is the set of inputs on which the third iteration is achievable by successively choosing $\bA_i$ and $\bA_j$ in the first and second iterations. $D_2\triangleq D_{(11)}\cup D_{(12)}\cup D_{(21)}\cup D_{(22)}$ is the set of inputs which allow at least three iterations. By simple calculation, we obtain $D_{(11)}=\{(0,0)\}$, $D_{(12)}=\{(0,0),(-1,1)\}$, $D_{(21)}=\{(0,0),(1,-1)\}$, $D_{(22)}=\{(1,-1)\}$, and $D_2=\{(0,0),(1,-1),(-1,1)\}$.
\item Observe that $D_1=D_2$. Our results obtained in this paper guarantee that $D=D_1=\{(0,0),(1,-1),(-1,1)\}$, namely, $D_1$ is actually the set of inputs which make the program possibly diverging.
\item Note that $V\cap D=\{(1,-1)\}\neq\emptyset$, yielding that the program is diverging on input $(x_0,y_0)=(1,-1)$.
\end{enumerate}
\oomit{\begin{figure}
\centering
\begin{tikzpicture}[->,>=stealth',shorten >=1pt,auto,node distance=1.2cm,semithick]
\node[state,rectangle](q1){$(1,-1)$};
\node[state,rectangle,white](q3)[below of=q1]{};
\node[state,rectangle](q2)[left of=q3]{$(0,0)$};
\node[state,rectangle](q4)[right of=q3]{$(1,1)$};

\node[state,rectangle](q6) at (-.6,-2.7) {$(0,0)$};
\node[state,rectangle](q5)[left of=q6]{$(-1,1)$};

\node[state,rectangle,white](q9) at (-2.35,-4.25) {$0$};
\node[state,rectangle,white](q10)[right of=q9]{$2$};
\node[state,rectangle,white](q11) at (-1.1,-4.25) {$0$};
\node[state,rectangle,white](q12)[right of=q11]{$2$};

\draw[->](q1)edge[swap](q2);
\draw[->](q1)edge[swap](q4);
\draw[->](q2)edge[swap](q5);
\draw[->](q2)edge[swap](q6);

\draw[->](q5)edge[swap,dotted](q9);
\draw[->](q5)edge[swap,dotted](q10);
\draw[->](q6)edge[swap,dotted](q11);
\draw[->](q6)edge[swap,dotted](q12);

\node at (-.75,-.45) {$1$};
\node at (.75,-.45) {$2$};
\node at (-1.7,-1.85) {$1$};
\node at (-0.7,-1.85) {$2$};
\node at (-2.25,-3.3) {$1$};
\node at (-1.3,-3.3) {$2$};
\node at (-0.95,-3.3) {$1$};
\node at (-0.15,-3.3) {$2$};
\end{tikzpicture}
\vspace*{-0.7cm}
\caption{\small The execution tree of \cref{exm:running} on input $(1,-1)$}\label{fig:exe-path-1}
\end{figure} }
%

The upper bound given in \cref{sec:bound} on the length of polynomial ascending chains (i.e., Hilbert ascending chains of polynomial ideals) guarantees that we only need to symbolically execute the loop for finitely many steps and check whether the given input is a solution to the ideal formed by some path in the induced finite execution tree.

We further remark that, for the above overview example, termination/nontermination of the loop cannot be concluded by existing techniques based on under/over-approximating the set $D$ of non-terminating inputs~\cite{LXZZ14}. The intuition is that for any input in $D$ there exist both a finite and an infinite path in the execution tree, leading to an empty set as the under-approximation of $D$ and a non-empty set as the over-approximation of $D$ (if achievable), and therefore no information about termination can be inferred from these approximations. See~\cite{LXZZ14} for a detailed explanation.

\oomit{
\section{A running example} \label{sec:nutshell}
Consider the following polynomial program (denoted by \verb|running|):
\begin{exm}\label{exm:running}
\begin{equation}\begin{split}
&(x,y) \coloneqq (x_0,y_0);\\
&\textsc{ while}\ (x+y=0)\ \textsc{do}\\
&\quad \textsc{if ? then}\ (x,y) \coloneqq (y^2,2x+y);\\
&\quad \textsc{else}\ (x,y) \coloneqq (2x^2+y-1,x+2y+1);\\
&\textsc{ end \ while}
\end{split}\end{equation}
Here ``\textsc{?}'' means that the condition takes a random Boolean value, and thus in each iteration 
one of the two assignments is \textit{nondeterministically} selected. 
What we concern is to decide if or not for any initial value $(x_0,y_0)$ in a given set $V=\{(x,y)\mid x^2+y=0\}$, the program would always terminate in a finite number of iterations.
\end{exm}
For simplicity, the polynomial of the loop guard is denoted as $G(x,y)=x+y$, and the two polynomial vectors of the assignments as $\bA_1(x,y)=(y^2,2x+y)$ and $\bA_2(x,y)=(2x^2+y-1,x+2y+1)$. Our approach is to compute the set $D$ of all possible initial values of $(x_0,y_0)$ for which the program may not terminate. Thus,  the termination problem of the program on
the set $X$ of inputs  is easily obtained by checking whether $X\cap D=\emptyset$. The procedure is outlined as follows:
\begin{enumerate}
\item Consider the equation $G(x,y)=x+y=0$, and write the set of its solutions as $D_0$. Thus, $(x_0,y_0)\in D_0$ means that the body
    of the loop should be executed once at least w.r.t.~the input. 
\item Denote by $D_{(1)}$ and $D_{(2)}$ the solution sets of equations:
\begin{equation*}\begin{split}
&\left\{
   \begin{aligned}
   &G(x,y)=x+y=0\\
   &G(\bA_1(x,y))=y^2+(2x+y)=0
   \end{aligned}\right.\\
{\rm and}&\\
&\left\{
   \begin{aligned}
   &G(x,y)=x+y=0\\
   &G(\bA_2(x,y))=(2x^2+y-1)+(x+2y+1)=0
   \end{aligned}\right.
\end{split}\end{equation*}
respectively. So,  $(x_0,y_0)\in D_{(i)}\ (i=1,2)$ means that the loop body may be executed
twice at least by correspondingly choosing $\bA_i$ to be the assignment in the first iteration. So $(x_0,y_0)\in D_1\triangleq D_{(1)}\cup D_{(2)}$ allows at least two iterations in the execution. It is easy to calculate that $D_{(1)}=\{(0,0),(-1,1)\}$, $D_{(2)}=\{(0,0),(1,-1)\}$, and so $D_1=\{(0,0),(1,-1),(-1,1)\}$.
\item Similarly, the solution set $D_{(ij)}$ of equation \begin{equation*}
\left\{
   \begin{aligned}
   &G(x,y)=0\\
   &G(\bA_i(x,y))=0\\
   &G(\bA_j(\bA_i(x,y)))=0
   \end{aligned}\right.
\end{equation*}
is the set of inputs for which the third iteration is achievable by successively choosing $\bA_i$ and $\bA_j$ in the first and the second iterations. $D_2\triangleq D_{(11)}\cup D_{(12)}\cup D_{(21)}\cup D_{(22)}$ is the set of inputs which allow at least three iterations. By simple calculation, we obtain that $D_{(11)}=\{(0,0)\}$, $D_{(12)}=\{(0,0),(-1,1)\}$, $D_{(21)}=\{(0,0),(1,-1)\}$, $D_{(22)}=\{(1,-1)\}$, and $D_2=\{(0,0),(1,-1),(-1,1)\}$.
\item Note that $D_1=D_2$. Our results reported in this paper guarantee that $D=D_1=\{(0,0),(1,-1),(-1,1)\}$, namely, $D_1$ is actually the set of inputs which make the program possibly nonterminating.
\item Observe that $V\cap D=\{(1,-1)\}\neq\emptyset$. So the program is nonterminating on input $(x_0,y_0)=(1,-1)$.
\end{enumerate}
\oomit{\begin{figure}
\centering
\begin{tikzpicture}[->,>=stealth',shorten >=1pt,auto,node distance=1.2cm,semithick]
\node[state,rectangle](q1){$(1,-1)$};
\node[state,rectangle,white](q3)[below of=q1]{};
\node[state,rectangle](q2)[left of=q3]{$(0,0)$};
\node[state,rectangle](q4)[right of=q3]{$(1,1)$};

\node[state,rectangle](q6) at (-.6,-2.7) {$(0,0)$};
\node[state,rectangle](q5)[left of=q6]{$(-1,1)$};

\node[state,rectangle,white](q9) at (-2.35,-4.25) {$0$};
\node[state,rectangle,white](q10)[right of=q9]{$2$};
\node[state,rectangle,white](q11) at (-1.1,-4.25) {$0$};
\node[state,rectangle,white](q12)[right of=q11]{$2$};

\draw[->](q1)edge[swap](q2);
\draw[->](q1)edge[swap](q4);
\draw[->](q2)edge[swap](q5);
\draw[->](q2)edge[swap](q6);

\draw[->](q5)edge[swap,dotted](q9);
\draw[->](q5)edge[swap,dotted](q10);
\draw[->](q6)edge[swap,dotted](q11);
\draw[->](q6)edge[swap,dotted](q12);

\node at (-.75,-.45) {$1$};
\node at (.75,-.45) {$2$};
\node at (-1.7,-1.85) {$1$};
\node at (-0.7,-1.85) {$2$};
\node at (-2.25,-3.3) {$1$};
\node at (-1.3,-3.3) {$2$};
\node at (-0.95,-3.3) {$1$};
\node at (-0.15,-3.3) {$2$};
\end{tikzpicture}
\vspace*{-0.7cm}
\caption{\small The execution tree of \cref{exm:running} on input $(1,-1)$}\label{fig:exe-path-1}
\end{figure} }
}

%% file: preliminaries.tex
\section{Preliminaries}\label{sec:pre}
\subsection{Polynomial rings and ideals}
A monomial of ${\bx}$ is of the form  $\bx^{\bal}\triangleq x_1^{\alpha_1} x_2^{\alpha_2}\cdots x_d^{\alpha_d},$
where ${\bal}=(\alpha_1,\alpha_2,\ldots,\alpha_d)\in\mathbb N^d$ is a vector of natural numbers, and
$\deg(\bx^{\bal}) \triangleq \sum_{i=1}^d \alpha_i$ is called the \emph{degree} of $\bx^{\alpha}$. A polynomial $f$ of ${\bx}$ is a linear combination of a finite number of monomials over a field $\bbK$ (either $\mathbb{Q}$, $\mathbb{R}$ or $\mathbb{C}$ in this paper), i.e., 
$f=\sum_{i=1}^m\lambda_i\bx^{\bal_i},$
where $m$ is the number of distinct monomials of $f$ and $\lambda_i\in\bbK$ is the nonzero coefficient of $\bx^{\bal_i}$, for each $i$. The degree of $f(\bx)$ is defined as  $\deg(f) \triangleq \max\{\deg(\bx^{\bal_i})\mid i=1,2,\ldots,m\}$.
Denote by $\mathcal{M}[\bx]$ the set of monomials of $\bx$ and $\bbK[\bx]$ the polynomial ring of $\bx$ over $\bbK$. The degree of a finite set $X\subseteq\bbK[\bx]$ is defined as $\deg(X) \triangleq \max\{\deg(f)\mid f\in X\}$.

We introduce the \emph{lexicographic order} over monomials: $\bx^{\bal}\prec\bx^{\pmb{\beta}}$ if there exists $1\leq i\leq d$ such that
$\alpha_i<\beta_i$ and $\alpha_j=\beta_j$ for all $1\leq j < i$. For every polynomial $f\in\bbK[\bx]$ we write its leading monomial (i.e., the greatest monomial under $\prec $) as $\lm(f)$. For any $n\in\bbN$, a set of monomials $M$ is called \textit{$n$-compressed}, if for any $m\in M$ with $\deg(m)=n$, $\{m'\mid\deg(m')=n,m'\prec m\}\subseteq M.$

\begin{defn}[Polynomial Ideal]\noindent
\begin{enumerate}\item A nonempty subset $I\subseteq\bbK[\bx]$ is called an \emph{ideal} if $f, g\in I\implies f+g\in I$, and $f\in I, h\in\bbK[\bx]\implies f h\in I$.
\item Let $P=\{f_1,f_2,\ldots,f_n\}$ be a nonempty subset of $\bbK[\bx]$. The \emph{ideal generated by $P$} is defined as $\langle P\rangle\triangleq\Big\{\sum_{i=1}^n f_i h_i\mid f_i\in P, h_i\in \bbK[\bx]\Big\}.$ $P$ is called a \emph{basis} of $ \langle P\rangle$.
\item The \emph{product} of two ideals $I$ and $J$ is defined as
\[I\times J \triangleq \langle\{f g\mid f\in I, g\in J\}\rangle.\]
\end{enumerate}\end{defn}
The ideal generated by $P$ is actually the minimal one that contains $P$. When $P=\{f_1,f_2,\ldots,f_n\}$ is a finite set, we simply write $\langle P\rangle$ as $\langle f_1,f_2,\ldots,f_n\rangle$. Given two polynomial sets $P$ and $Q$, we define $P\cdot Q\triangleq\{f g\mid f\in P, g\in Q\}$. Obviously, $\langle P\rangle\times\langle Q\rangle=\langle P\cdot Q\rangle$.
%

Hilbert's basis theorem~\cite[Chapter~7]{Atiyah1969} implies that the polynomial ring $\bbK[\bx]$ is a Noetherian ring, that is,
\begin{thm}[Ascending Chain Condition]\label{thm:asc-chain}For any ascending chain of ideals
	\[I_1\subseteq I_2\subseteq \cdots\subseteq I_n \subseteq \cdots\]
	of\, $\bbK[\bx]$, there exists $N\in\mathbb{N}$ such that $I_n=I_N$ for all $n\geq N$.\end{thm}

An equivalent statement of \cref{thm:asc-chain} is that every ideal $I\subseteq \bbK[\bx]$ is \emph{finitely-generated}, that is, $I=\langle f_1,f_2,\ldots,f_n\rangle$ for some $f_1,f_2,\ldots,f_n\in \bbK[\bx]$. Thus, we can define the degree of an ideal $I$ as $\gdeg(I)=\min\{\deg(P)\mid P\ \text{is a basis of}\ I\}$.

Note that an ideal may have different bases. However, using Buchberger's algorithm~\cite{buchberger1965algorithmus} under a fixed monomial ordering, a unique (reduced) Gr\"{o}bner basis of $I$, denoted $\GB(I)$, can be computed from any other basis of $I$. We simply write $\GB(\langle P\rangle)$ as $\GB(P)$ for any basis $P$. An important property of Gr\"{o}bner bases is that the remainder of any polynomial $f$ on division by $\GB(P)$, written as $\Rem(f,\GB(P))$, satisfies
\[f\in\langle P\rangle\iff\Rem(f,\GB(P))=0.\]

\subsection{Algebraic sets and varieties}
 A polynomial $f\in\bbK[\bx]$ can be regarded as a function from the affine space $\bbK^d$ to $\bbK$. The set of zeros of a polynomial set $P\subset\bbK[\bx]$ can be defined as $Z(P)\triangleq\{\bx\in\bbK^d\mid\forall f\in P\colon f(\bx)=0\}$. It is trivial to verify that $Z(P)=Z(\langle P\rangle)$.
\begin{defn}[Algebraic Set and Variety] A subset $X\subseteq\bbK^d$\begin{enumerate}
\item is \emph{algebraic}, if there exists $P\subseteq\bbK[\bx]$ such that $X=Z(P)$, and $P$ is called a set of \emph{generating polynomials of $X$};
\item is \emph{reducible}, if it has two proper algebraic subsets $X_1$ and $X_2$ such that $X=X_1\cup X_2$; otherwise it is called \emph{irreducible};
\item is a \emph{variety}, if it is a nonempty irreducible algebraic set.
\end{enumerate}\end{defn}
 The following properties on algebraic sets and varieties hold: the union of two algebraic sets is an algebraic set, and the intersection of any family of algebraic sets is algebraic; suppose $X_1,X_2,\ldots,X_n$ are algebraic sets and $X$ is a variety, then
\[
X\subseteq X_1\cup\cdots\cup X_n\implies \exists k \in \{1,\ldots,n\}\colon X\subseteq X_k.
\]
An algebraic set is usually represented by a (not necessarily unique) set of generating polynomials. 
By defining
\[I(X)\triangleq\{f\in\bbK[\bx]\mid\forall \bx\in X\colon f(\bx)=0\}\]
for any $X\subseteq \bbK^d$, one can verify that $I(Z(P))$ is the maximal set that generates $Z(P)$. Hence, any algebraic set $X$ can be identified by the ideal $I(X)$. The membership $f\in I(Z(P))$ for any polynomial $f$ and any finite set $P=\{f_1,\ldots,f_n\}$ of polynomials is equivalent to the unsatisfiability of $f_1=0\wedge f_2=0\wedge\cdots\wedge f_n=0\wedge f\neq0$, which is decidable due to Tarski~\cite{Tarski51}.

Additionally, considering the fact that $X_1\subset X_2\iff I(X_1)\supset I(X_2)$ for any two algebraic sets $X_1$ and $X_2$, it follows immediately from \cref{thm:asc-chain} that
\begin{thm}[Descending Chain Condition]\label{thm:des-chain}For any descending chain of algebraic sets
\[X_1\supseteq X_2\supseteq\cdots\supseteq X_n\supseteq\cdots\]
of\, $\bbK^d$, there exists $N\in\mathbb{N}$ such that $X_n=X_N$ for all $n\geq N$.\end{thm}

\subsection{Monomial ideals and Hilbert's functions}
An ideal $I$ is called monomial if it can be generated by a set of monomials. A monomial ideal always has a finite basis $\{m_1,m_2,\ldots,m_s\}$ of monomials (by Dickson's Lemma), and any monomial $m\in \langle m_1,m_2,\ldots,m_s\rangle$ should be a multiple of some $m_i$.

\begin{defn}[Hilbert's Function]
For a monomial ideal $I\subseteq\bbK[\bx]$, \emph{Hilbert's function} $H_I \colon \bbN\rightarrow\bbN$ is defined as
\[H_I(n) \triangleq \dim_\bbK\bbK_n[\bx]/I_n=\dim_\bbK\bbK_n[\bx]-\dim_\bbK I_n,\]
where $\bbK_n[\bx]$ is the set of homogeneous polynomials of degree $n$ and $I_n=I\cap\bbK_n[\bx]$, and both of them are linear spaces over $\bbK$; $\dim_\bbK$ denotes the dimension of the linear spaces over $\bbK$.
\end{defn}
Note that $\dim_\bbK\bbK_n[\bx]=(n)_{d-1}$ is the number of monomials of degree $n$, where
\[(n)_{d-1}\triangleq\binom{n+d-1}{d-1}=\frac{(n+d-1)!}{n!\,(d-1)!}.\]
$H_I(n)=0$ means that $I$ contains all monomials of degree at least $n$.

We use Macaulay's theorem~\cite{FSM27} to estimate the value of Hilbert's function $H_I$. To this end, we define a function $\Inc_k \colon \bbN\rightarrow\bbN$ for natural number $k\geq 1$ as follows. When $k$ is given, any number $n\in\bbN$ can be uniquely decomposed as
\[n=(n_1)_k+(n_2)_{k-1}+\cdots+(n_r)_{k-r+1}\triangleq(n_1,n_2,\ldots,n_r)_k,\]
where $0\leq r\leq k$ and $n_1\geq n_2\geq\cdots\geq n_r\geq 0$. In fact, $0=()_k$ with $r=0$; and for $n>0$, $n_1, n_2,\ldots,n_r$ are successively determined by
\begin{equation*}
\begin{split}
(n_1)_k&\leq n<(n_1+1)_k\\
(n_2)_{k-1}&\leq n-(n_1)_k<(n_2+1)_{k-1}\\
&\cdots
\end{split}
\end{equation*}
until $n=(n_1,n_2,\ldots,n_r)_k$ for some $r\geq 1$. Now we define \[\Inc_k((n_1,\ldots,n_r)_k)=(n_1,\ldots,n_r)_{k+1}.\]
For instance, $\Inc_k(0)=0$, and $\Inc_3(11)=\Inc_3((2,0)_3)=(2,0)_4=16$ (note that $11=(2,0)_3$).
\begin{thm}[Macaulay's Theorem~\cite{FSM27}]\label{thm:Macaulay}
For any monomial ideal $I\subseteq\bbK[\bx]$,
\[H_I(n+1)\leq\Inc_n(H_I(n)) \text{ for all } n \geq 1.\]
Moreover, $H_I(n+1)=\Inc_n(H_I(n))$ if $\gdeg(I)\leq n$ and a monomial basis of $I$ is $n$-compressed.
\end{thm}

%% file: bound.tex
\section{Upper bound on the length of polynomial ascending chains} \label{sec:bound}

In this section, we exploit the maximal length of polynomial ascending chains. Our approach consists of the following three steps (a sketch only, with definitions given later):
 \begin{itemize}
 \item[(i)] Reduce the computation of the bound on $f$-bounded polynomial ideal chains to computing the bound on $f$-generating sequences of monomials (cf. \cref{prop:monomial}), which is obtained by Moreno-Soc\'{\i}as's result \cite{GMS92}.
 \item[(ii)] Construct the longest $f$-generating sequence of monomials whose degrees are precisely determined (not just upper-bounded) by $f$, leveraging Hilbert's function and Macaulay's theorem.
 \item[(iii)] Prove that the bound on $f$-generating sequences of monomials is identical to the length of the sequence obtained in (ii) over an extended space (cf. \cref{thm:bound}). This is achieved by introducing an auxiliary variable.
 \end{itemize}

\begin{rem}
	The use of Hilbert's function and Macaulay's theorem is inspired by Moreno-Soc\'{\i}as's proof tactics in \cite{GMS92}. There are, however, crucial differences in the underlying techniques for constructing the longest $f$-generating sequence of monomials: we perform the construction in a necessarily extended space of monomials $\mathcal{M}[\bx,x_{d+1}]$ (Step (iii), with an auxiliary variable $x_{d+1}$); in contrast, the construction in~\cite{GMS92} works directly on the original monomial space $\mathcal{M}[\bx]$. We will show in \cref{exm:incorrect} that the constructed $f$-generating sequence of monomials in~\cite{GMS92} is in fact not the longest (revealing that Proposition 4.3 in~\cite{GMS92} is wrong). It will also be shown that such a spuriously ``maximal'' length may lead to unsoundness issues when being applied in termination analysis of MPPs.
\end{rem}

Consider a field $\bbK$ of characteristic $0$ (interpreted as $\mathbb{Q}$, $\mathbb{R}$ or $\mathbb{C}$ for different types of programs). Let $\bx=(x_1,x_2,\ldots,x_d)$ be a vector of variables and $\bbK[\bx]$ the polynomial ring of $\bx$ over $\bbK$.
\begin{defn}[$f$-Bounded Chain of Polynomial Ideals]
	For any monotonically increasing function $f \colon \bbN\rightarrow\bbN$ (namely, $f(x)\leq f(y)$ for all $x\leq y$), an ascending chain $I_1\subseteq I_2\subseteq\cdots\subseteq I_n$ of polynomial ideals of $\bbK[\bx]$ is called \emph{$f$-bounded}, if $\gdeg(I_i)\leq f(i)$ for all $i\geq 1$. Let $L(d,f)$ denote the maximal length of all strictly ascending chains of $\bbK[\bx]$ that are $f$-bounded. Here, $d$ is the dimension of $\bx$.
\end{defn}
\begin{rems}\noindent
\begin{enumerate}\item The condition of $f$-boundedness is necessary to define the maximal length, as the length of a chain with unbounded degrees could be arbitrarily large. For instance, the length of $\langle x^n\rangle\subset\langle x^{n-1}\rangle\subset\cdots\subset\langle 1\rangle$ may be arbitrarily large if $n$ is unbounded.
\item For ease of discussion, the function $f$ is assumed to be increasing without loss of generality. In fact, for a general $f$, consider the increasing function $F \colon \bbN\rightarrow\bbN$ with $F(n)\triangleq\max\{f(1),f(2),\ldots,f(n)\}$. Then an $f$-bounded chain is always an $F$-bounded chain since $f(n)\leq F(n)$ for all $n$. Hence, $L(d,f)\leq L(d,F)$ and $L(d,F)$ is an upper bound on the lengths of the chains.
\end{enumerate}
\end{rems}

Our aim is to compute $L(d,f)$ in terms of the number $d$ of variables and function $f$. To this end, we particularly consider ascending chains of a special form.
\begin{defn}[$f$-Generating Sequence of Monomials]\label{def:f-gener}
	For a function $f \colon \bbN\rightarrow\bbN$, a finite sequence of monomials $m_1,m_2,\ldots,m_n\in\mathcal{M}[\bx]$ is called \emph{$f$-generating}, if $\deg(m_i)\leq f(i)$ and $m_{i+1}\notin\langle m_1,\ldots,m_i\rangle$ for all $i\geq 1$.
\end{defn}
Then an $f$-generating monomial sequence $m_1,m_2,\ldots,m_n$ generates a strictly ascending chain of monomial ideals $I_1\subset I_2\subset\cdots\subset I_n$ satisfying $\gdeg(I_i)\leq f(i)$, where $I_i \triangleq \langle m_1,m_2,\ldots,m_i\rangle$.
Moreno-Soc\'{\i}as proved in \cite{GMS92} that in order to compute $L(d,f)$, it suffices to consider the ascending chains that are generated by $f$-generating monomial sequences. Namely,
\footnote{\Cref{prop:monomial} is valid, as it only states that $L(d,f)$ is the largest number of monomials, yet without addressing how to construct $L(d,f)$, which is essentially the flawed part in~\cite{GMS92}.}
\begin{prop}[\cite{GMS92}]\label{prop:monomial}$L(d,f)$ is exactly the largest number of monomials of $f$-generating sequences in $\mathcal{M}[\bx]$.\end{prop}

Hence, in the rest of this section, we construct the longest chain of this form. We start with a special case where the degrees of polynomial ideals are not just bounded but completely determined by a function $f$. Then we reduce the general case to this special one.

\oomit{\begin{rem} In \cite{GMS92}, G.M. Soc\'{\i}as tried to compute $L(d,f)$ using Macaulay's theorem (i.e., \cref{thm:Macaulay}), but his final result is incorrect, because his construction of the longest $f$-generating sequence is wrong (a counter-example is
given in \cref{exm:incorrect}). In what follows we will propose the correct one.\end{rem} }

\subsection{The longest chain of specified degrees}
Consider a special type of $f$-generating sequences $m_1,\ldots,m_n$ such that
\begin{equation}\label{equ:special}\forall i\in\{1,\ldots,n\}\colon \deg(m_i)=f(i).\end{equation}

We inductively construct an $f$-generating sequence of monomials $\hat{m}_1,\hat{m}_2,\ldots,\hat{m}_N$ as follows: Initially define $\hat{m}_1=x_1^{f(1)}$; suppose $\hat{m}_1,\hat{m}_2,\ldots,\hat{m}_n$ are defined for some $n\geq 1$, then let
\begin{equation}\label{equ:hatm}\hat{m}_{n+1}\triangleq \min\{m \mid \deg(m)=f(n+1), m\notin \langle \hat{m}_1,\ldots,\hat{m}_n\rangle\}
\end{equation}
until $\{m \mid \deg(m)=f(N+1)\}\subseteq\langle \hat{m}_1,\ldots,\hat{m}_N\rangle$ holds for some $N$. Here $\min$ is taken with respect to the lexicographical ordering over monomials.

Obviously, this sequence satisfies \cref{equ:special}. It follows immediately from \cref{equ:hatm} that the corresponding ideal $\hat{I}_n\triangleq\langle\hat{m}_1,\hat{m}_2,\ldots,\hat{m}_n\rangle$ is $f(n)$-compressed for every $n=1,2,\ldots,N$, and $H_{\hat{I}_N}(f(N+1))=0$ holds by the definition of Hilbert's function.

\begin{exm}\label{exm:4382}
For $d=3$ and $f(n)=2\times3^{n-1}$, we have
\begin{multline*}
\{\hat{m}_1,\hat{m}_2,\ldots,\hat{m}_{4382}\}=\{x^2,xy^5,xy^4z^{13},xy^3z^{50},xy^2z^{159},\\
xyz^{484}, xz^{1457},y^{4374},y^{4373}z^{8749},\ldots,y^0z^{2\cdot3^{4381}}\}.
\end{multline*}
Then, $N=4382$ in this case.
\end{exm}

Now we prove that the constructed sequence has the largest number of monomials among all $f$-generating sequences that satisfy \cref{equ:special}.
\begin{lem}\label{lem:sequence-bound}
	If $m_1,m_2,\ldots,m_n$ is an $f$-generating sequence that satisfies \cref{equ:special}, then $n\leq N$.
\end{lem}

\oomit{
	\begin{proof}[Proof (of \cref{lem:sequence-bound})]
		Write $I_i=\langle m_1,\ldots,m_i\rangle$ for all $i\in{1,2,\ldots,n}$, then it suffices to prove that $H_{I_i}(j)\leq H_{\hat{I}_i}(j)$ for all $i,j\geq 1$. This can be done by induction on $i+j$ and the details are omitted here.
	\end{proof}
}
\begin{proof}
	We prove by contradiction.  Suppose $n\geq N+1$. 
	Let $I_i=\langle m_1,\ldots,m_i\rangle$ for all $i\in \{1,2,\ldots,n\}$.
	For simplicity, we define $H_i(j)\triangleq H_{I_i}(j)$ and $\hat{H}_{i}(j)\triangleq H_{\hat{I}_i}(j)$, for all $i,j\geq 1$. Observe that\begin{equation}\label{equ:H_exp}
	H_i(j)\left\{
	\begin{aligned}
	&=H_{i-1}(j),\quad j<f(i);\\
	&=H_{i-1}(j)-1,\quad j=f(i);\\
	&\leq \Inc_{j-1}H_i(j-1),\quad j>f(i).
	\end{aligned}\right.
	\end{equation}
	The first two equalities follow directly from the definition of $I_i$, and the third inequality from \cref{thm:Macaulay}. Similarly, we have\begin{equation}\label{equ:hatH_exp}
	\hat{H}_i(j)\left\{
	\begin{aligned}
	&=\hat{H}_{i-1}(j),\quad j<f(i);\\
	&=\hat{H}_{i-1}(j)-1,\quad j=f(i);\\
	&=\Inc_{j-1}\hat{H}_i(j-1),\quad j>f(i).
	\end{aligned}\right.
	\end{equation}
	Here, the third one becomes an equality since $\hat{I_i}$ is $f(i)$-compressed and $\gdeg(\hat{I_i})=f(i)\leq j-1$, and thus the conditions for the equality in \cref{thm:Macaulay} are satisfied. On the other hand, we observe that $H_1(f(1))=\hat{H}_1(f(1))=(f(1))_{d-1}-1$. It then can be proved inductively from \cref{equ:H_exp,equ:hatH_exp} that
	\begin{equation*}\begin{split}
	H_i(j)&=H_{i-1}(j)\leq \hat{H}_{i-1}(j)=\hat{H}_i(j),\quad  j<f(i);\\
	H_i(j)&=H_{i-1}(j)-1\leq \hat{H}_{i-1}(j)-1=\hat{H}_i(j),\quad j=f(i);\\
	H_i(j)&\leq \Inc_{j-1}H_i(j-1)\leq\Inc_{j-1}\hat{H}_i(j)=\hat{H}_i(j),\quad j>f(i).
	\end{split}\end{equation*}
	Here, the fact that $r \leq t\implies\Inc_j(r)\leq \Inc_j(t)$ is applied. Hence we have proved that $H_i(j)\leq\hat{H}_i(j)$ for all $i\geq 1$ and $j\geq 1$. Then $H_N(f(N+1))\leq \hat{H}_N(f(N+1))=0$. We have $m_{N+1}\in\{m \mid \deg(m)=f(N+1)\}\subseteq\langle \hat{m}_1,\ldots,\hat{m}_n\rangle$, which contradicts \cref{equ:hatm}.
\end{proof}

We consider next how to compute the maximal length $N$ from the longest chain $\hat{m}_1,\ldots,\hat{m}_N$ as constructed in \cref{equ:hatm}. To this end, we define $\Omega(d,f,t)$ to be the number $k$ such that the $k$-th monomial of the chain is
\[
\hat{m}_k=x_1^{f(1)-t}x_d^{f(k)-f(1)+t}.
\]
Then from this definition $N=\Omega(d,f,f(1))$. The procedure to compute $\Omega(d,f,t)$ is described by the following theorem.

\begin{thm}\label{thm:Omega}
Given $d,t\in\bbN$ and an increasing function $f \colon \bbN\rightarrow\bbN$, $\Omega(d,f,t)$ can be recursively computed as follows.
\begin{enumerate}
\item $\Omega(1,f,t)=1$ and $\Omega(d,f,0)=1$, for any $d\geq 1$, $f$ and $t$.
\item Let $n_t\triangleq \Omega(d,f,t)$, recursively calculated by
\[n_t=n_{t-1}+\Omega(d-1,f_{n_{t-1},t-f(1)},f_{n_{t-1},t-f(1)}(1)).\]
Here $f_{m,r}$ is a function defined as $f_{m,r}(n)\triangleq f(m+n)+r$.
\end{enumerate}
\end{thm}

\begin{proof}
	By the construction of the chain $\{\hat{m}_k\}_k.\!\!$
\end{proof}

For instance, let $f(n)=2\times 3^{n-1}$, then $\Omega(2,f,2)=4382$ by \cref{thm:Omega}, which is exactly the number of monomials in \cref{exm:4382}.

\subsection{Reduction from the general case}
Now we drop the restriction in \cref{equ:special} and consider the length of a general $f$-generating sequence of monomials $m_1,\ldots,m_n$ in $\mathcal{M}[\bx]$. The idea is to reduce this general case to the  special case. To that end, we introduce a new variable $x_{d+1}$ (for which the lexicographic order is based on the order among variables  $x_1\succ \cdots \succ x_d\succ x_{d+1}$), and construct for each $m_i$ $(i=1,\ldots,n)$ a monomial $\tilde{m}_i\in\mathcal{M}[\bx,x_{d+1}]$ such that $\tilde{m}_i=m_ix_{d+1}^{c_i}$, where $c_i=f(i)-\deg(m_i)\in\bbN$. Hence $\deg(\tilde{m}_i)=f(i)$, and thus \cref{equ:special} is satisfied by $\tilde{m}_1,\ldots,\tilde{m}_n$. Furthermore,

\begin{prop}\label{prop:sequence-K}
	$\tilde{m}_1,\ldots,\tilde{m}_n$ is an $f$-generating sequence of $\bbK[\bx,x_{d+1}]$.
\end{prop}

\begin{proof}
	It suffices to prove that for every $i=1,2,\ldots,n-1$, $\tilde{m}_{i+1}\notin\langle\tilde{m}_1,\ldots,\tilde{m}_{i}\rangle$. In fact, if this is not the case then $\tilde{m}_{i+1}$ should be a multiple of some $\tilde{m}_j$, with $j\leq i$. It implies that $m_{i+1}$ is a multiple of $m_j$, which contradicts $m_{i+1}\notin\{m_1,\ldots,m_i\}$.
\end{proof}

It then follows immediately from the definition of $\Omega$ that $n\leq\Omega(d+1,f,f(1))$. Since $m_1,\ldots,m_n$ is arbitrarily chosen, we obtain $L(d,f)\leq\Omega(d+1,f,f(1))$ by \cref{prop:monomial}. Conversely, we also show that $\Omega(d+1,f,f(1))\leq L(d,f)$. Consider the sequence $\hat{m}_1,\ldots,\hat{m}_N$ of $\bbK[\bx,x_{d+1}]$ defined as in \cref{equ:hatm}. Here $N=\Omega(d+1,f,f(1))$. By letting $x_{d+1}=1$, the obtained sequence $m_1^\prime,\ldots,m_N^\prime$ of $\bbK[\bx]$ satisfies

\begin{prop}\label{prop:sequence}
	$m_1^\prime,\ldots,m_n^\prime$ is an $f$-generating sequence.
\end{prop}

\begin{proof}
	Observe that $\deg(m_i^\prime)\leq\deg(\tilde{m}_i)=f(i)$ for all $i$. Then it remains to prove that for all $i$, $m_i^\prime\notin\{m_1^\prime,\ldots,m_{i-1}^\prime\}$. Assume that $m_i^\prime$ is a multiple of some $m_j^\prime$, with $j<i$. Let $\tilde{m}_i=m_i^\prime x_{d+1}^\alpha$ and $\tilde{m}_j=m_j^\prime x_{d+1}^\beta$. Then $\alpha<\beta$, otherwise $\tilde{m}_i$ is a multiple of $\tilde{m}_j$. We also have $f(j)<f(i)$, otherwise, $f(j)=f(i)$ and thus $\tilde{m}_i\prec\tilde{m}_j$, which contradicts $j<i$.
	
	Note that $\deg(m_j^\prime)=f(j)-\beta< f(j)-\alpha<f(i)-\alpha=\deg(m_i^\prime)$. Then we can find some monomial $m'\in\mathcal{M}[\bx]$ such that $\deg(m')=f(j)-\alpha$, and $m'$ is a multiple of $m_j^\prime$ and also a factor of $m_i^\prime$. Let $\tilde{m}=m'x_{d+1}^\alpha\in\mathcal{M}[\bx,x_{d+1}]$, then $\deg(\tilde{m})=f(j)=\deg(\tilde{m}_j)$ and $\tilde{m}\prec \tilde{m}_j$. Thus $\tilde{m}$ is also in the sequence $\tilde{m}_1,\ldots,\tilde{m}_n$. This contradicts that $\tilde{m}_i$ is a multiple of $\tilde{m}$.
\end{proof}

$\Omega(d+1,f,f(1))=N\leq L(d,f)$ follows immediately from this result. Therefore, we obtain the following fact.
\begin{thm}\label{thm:bound} $L(d,f)=\Omega(d+1,f,f(1))$.\end{thm}

\begin{exm}\label{exm:incorrect}
For $d=2 ,f(i)=1+i$, the monomial set of the longest monomial ascending chain is
\[\{ m_1,m_2,\ldots,m_{11}\}=\{x^2,xy^2,xy,x,y^6,y^5,y^4,y^3,y^2,y,1\},\]
and $L(2,f) = 11$.
\end{exm}
However, according to Moreno-Soc\'{i}as's approach \cite{GMS92}, the monomial set is
\[
\{x^2,xy^2,y^4,y^3,xy,y^2,x,y,1\},
\]
and thus $L(2,f) = 9$, which is evidently smaller (than $11$) and can be proven not to be the maximal length of all the ascending chains (by applying \cref{def:f-gener}). We remark that such a spurious bound on the length may lead to crucial consequences when being applied in termination analysis of MPPs: a non-terminating MPP can be erroneously concluded as being terminating because of not reaching a sufficient depth in its execution tree (which is captured by the maximal length of the ascending chains). Thus termination analysis based on this bound may give unsound results.

\subsection{Complexity of \texorpdfstring{$L(d,f)$}{L(d,f)}}
To show that $L(d,f)$ is not primitive recursive, we introduce the Ackermann function:
\begin{defn}[Ackermann Operator]The \emph{Ackermann operator} $\Uparrow^d$ for natural numbers $d,x,y$ is defined by
{\reqnomode
\begin{align*}
x\Uparrow^0 y & = 1+xy\\
x\Uparrow^d 0 & = 1 \tag{$d > 0$}\\
x\Uparrow^d y & = x\Uparrow^{d-1} (x\Uparrow^d y-1) \tag{$d>0, y>0$}
\end{align*}
}
\end{defn}
For instance, one can easily compute that $x\Uparrow^d 1=1+x$ and
\[x\Uparrow^1 y=1+x+x^2+\cdots+x^y.\]
A basic property of the Ackermann function is that it is not primitive recursive. Now we use it to discuss the complexity of $L(d,f)$.

\begin{prop}\label{prop:Ackermann}
	For two positive integers $a, b$, and $f(n)=a(n-1)+b$ with $n \ge 1$,
	\begin{equation}\label{equ:linear-length}
		L(d,f)=\frac{1}{a}\left((a+1)\Uparrow^{d-1}(b+1)-b-2\right).
	\end{equation}
\end{prop}

\begin{proof}
	For the function $f(n)=a(n-1)+b$, let $\Phi(d,a,b)\triangleq a\Omega(d,f,f(1))+b+1$, where $\Omega(d,f,t)$ is defined in \cref{thm:Omega}. Since $f(1)=a(1-1)+b=b$, we have
	$\Phi(d,a,b) = a\Omega(d,f,b)+b+1$.
	Further by \cref{thm:Omega},  we have
	$\Phi(1,a,b)=1+a+b$, $\Phi(d,a,0)=1+a$, and $\Phi(d,a,b)=\Phi(d-1,a,\Phi(d,a,b-1))$.
	It follows that
	\[\Phi(2,a,b)=(1+a)(1+b)=(1+a)\Uparrow^0(1+b)-1,\]
	and it can be proved by induction on $d$ that for all $d\geq 2$,
	\[\Phi(d,a,b)=(1+a)\Uparrow^{d-2}(1+b)-1.\]
	Then \cref{equ:linear-length} follows immediately from \cref{thm:bound}.
\end{proof}

In \cref{prop:Ackermann}, $L(d,f)$ is a primitive recursive function on $a$ and $b$ for a fixed $d$, yet becomes Ackermannian when $d$ is not fixed (for general $f$). This observation coincides with Figueira et al.'s conclusion in~\cite{FFSS11} that $L(d,f_{t,0})$, a function on $t$, is at level $\Gamma_{\lambda+k-1}$ of the fast-growing hierarchy for $f$ at level $\lambda$. Their results answered Seidenberg's second question on non-primitive recursiveness by giving a lower bound on $L(d,f)$. In contrast, our approach gives an explicit recursive function (which is Ackermannian) as the maximal length $L(d,f)$, thereby an answer also to Seidenberg's request 
for an explicit expression of the bound. 
Despite that the computation of the bound is not necessary to show decidability, an explicit expression for the bound, however, delivers insight on how many iterations a loop at most will take until termination, and hence is useful in building a decision procedure for termination.

\oomit{
\begin{rem}
The decidability of termination of MPPs is not a claim in~\cite{FFSS11}, though we observe that the generic theory therein could be applicable to MPPs and may potentially lead to a decidable result in principle. Nevertheless, this generic theory is not directly applicable in building a decision procedure for termination, due to the lack of an explicit expression of the bound. Therefore, despite that the computation of the bound is not necessary to show decidability, an explicit expression for the bound, however, delivers insight on how many iterations at most a loop will take until termination.
\end{rem}
}

%% file: mpps.tex
\section{Termination of MPPs with equality guards}\label{sec:MPP}
\subsection{Multi-path polynomial programs (MPPs)}
The family of polynomial programs considered in this section is formally defined as
\begin{defn}[MPPs with Equality Guards \cite{LXZZ14}]\label{def:mpp}
A \emph{multi-path polynomial program with equality guard} is of the form
\begin{equation}\label[mpp]{equ:mpp}
\textsc{ while} \quad (G(\bx)=0)\quad
  \left\{
   \begin{aligned}
   &\bx\coloneqq \bA_{1}(\bx);\\
   \parallel\quad &\bx\coloneqq \bA_{2}(\bx); \\
   &\vdots\\
   \parallel\quad &\bx\coloneqq \bA_{l-1}(\bx); \\
   \parallel\quad &\bx\coloneqq \bA_{l}(\bx);
   \end{aligned}\right\},
\end{equation}
where
\begin{enumerate}
    \item $\bx\in \bbK^d$ denotes the vector of program variables;
    \item $ G\in \bbK[\bx]$ is a polynomial and $G({\bx})=0$ is the loop condition;
    \item $ \bA_{i}\in \bbK^d[\bx]$ ($ 1\leq i\leq l$) are vectors of polynomials, describing the assignments to program variables in the loop body;
    \item $\parallel$ is interpreted as a nondeterministic choice between the $l$ assignments.
  \end{enumerate}
\end{defn}
\begin{rems}\noindent
\begin{enumerate}
\item \oomit{Whenever the set $\{\bx\in\bbK^d\mid\bx\ $ satisfies the loop guard $\}$ is algebraic, decidability of termination can be obtained using our method in the paper. Therefore,} The loop guard of \cref{equ:mpp} can be extended to a more general form $\bigvee_{i=1}^M\bigwedge_{j=1}^{N_i} G_{ij}(\bx)=0$. It is nevertheless essential to assume that inequalities will never occur in guards, otherwise the termination problem will become undecidable, even not semi-decidable~\cite{bradley:polynomial}.
\item The initial value of $\bx$ is not specified, and is assumed to be taken from $\mathbb{K}^{d}$. If the input $\bx$ is subject to
semi-algebraic constraints, our decidability result still holds according to \cite{Tarski51}.
\end{enumerate}
\end{rems}

\begin{exm}[\Verb!liu1!~\cite{LXZZ14}]
\label{exm1}
Consider the MPP
\begin{equation*}
\textsc{while} \quad (x^2+1-y=0)\quad
  \left\{
   \begin{aligned}
   (x,y) &\coloneqq (x,x^2y);  \\
   \parallel\quad (x,y) &\coloneqq (-x,y);  \\
    \end{aligned}
   \right\}.
\end{equation*}
We have $d=2$, $G(x,y)=x^2-y+1$, $\bA_1(x,y)=(x,x^2y)$ and $\bA_2(x,y)=(-x,y)$.
\end{exm}

\begin{exm}\label{exa:quantum}
Consider a nondeterministic quantum program~\cite{LY14} of the form
\begin{equation*}\label[mpp]{equ:quantum}
\textsc{while} \quad (\textit{Mea}[\rho]=0)\quad
  \left\{
   \begin{aligned}
   \rho &\coloneqq \mathcal{E}_1(\rho);  \\
   \parallel\quad \rho &\coloneqq \mathcal{E}_2(\rho);  \\
   &\vdots\\
   \parallel\quad \rho &\coloneqq \mathcal{E}_{l-1}(\rho); \\
   \parallel\quad \rho &\coloneqq \mathcal{E}_l(\rho);
    \end{aligned}
   \right\},
\end{equation*}
where
\begin{enumerate}
\item $\rho\in\mathbb{C}^{d^2}$ is a $d\times d$ density matrix.
\item $\textit{Mea}=\{M_0,M_1\}$ is a two-outcome quantum measurement, where $M_0$ and $M_1$ are $d\times d$ complex matrices.
\item $\mathcal{E}_i$ are quantum super-operators that perform linear transformations over $\mathbb{C}^{d^2}$.
\end{enumerate}
 In this example, $\bx=\rho$, $G(\bx)={\rm tr}(M_0\rho M_0^\dagger)$ and $\bA_i=\mathcal{E}_i$, 
 where ${\rm tr}(M)$ is the trace of a matrix $M$, and $M^\dagger=(M^\mathrm{T})^*$ is the complex conjugate of the transpose of $M$. Clearly, it is a multi-path linear program over $\mathbb{C}^{d^2}$. 
\end{exm}

In \cite{LY14}, the set of non-terminating inputs (diverging states, in their terminology) of the loop in \cref{exa:quantum} plays a key role in deciding the termination of nondeterministic quantum programs. We will show later in the experiments that our approach can be used to discover termination/nontermination proofs of such quantum programs.

\subsection{Execution of MPPs}
Given an input $\bx$, 
the behavior of \cref{equ:mpp} on $\bx$ is determined by the nondeterministic choices of $\bA_i\ (1\leq i\leq l)$, and consequently all the possible  executions form a tree.
\begin{defn}[Execution Tree \cite{LXZZ14}]
The \emph{execution tree} of \cref{equ:mpp} on input $\bx\in \bbK^d$ is defined inductively as follows:
 \begin{enumerate}
    \item The root is the input value of $\bx$.
    \item For any node $ \widetilde{\bx}$, it is a leaf node if $ G(\widetilde{\bx})\neq 0$; otherwise, $\widetilde{\bx}$ has $l$ children $\bA_{1}(\widetilde{\bx})$, $\bA_{2}(\widetilde{\bx}),\ldots,\bA_{l}(\widetilde{\bx})$, and there is a directed edge from $\widetilde{\bx}$ to $\bA_{i}(\widetilde{\bx})$, labeled by $i$, for $1\leq i\leq l$.
  \end{enumerate}
\end{defn}
\oomit{\begin{figure}
\centering
\begin{tikzpicture}[->,>=stealth',shorten >=1pt,auto,node distance=1.5cm,semithick]
\node[state,rectangle](q1){$\xx$};
\node[state,rectangle](q3)[below of=q1]{$\cdots$};
\node[state,rectangle](q2)[left of=q3]{${\bf A}_1(\xx)$};
\node[state,rectangle](q4)[right of=q3]{${\bf A}_{\ell}(\xx)$};
\node[state,rectangle](q6)[below of=q2]{$\cdots$};
\node[state,rectangle](q5)[left of=q6]{${\bf A}_1({\bf A}_1(\xx))$};
\node[state,rectangle](q7)[right of=q6]{${\bf A}_{\ell}({\bf A}_1(\xx))$};

\node[state,rectangle,white](q8)[below of=q4]{};
\node[state,rectangle,white](q9)[below of=q5]{};
\node[state,rectangle,white](q10)[below of=q7]{};

\draw[->](q1)edge[swap](q2); \draw[->](q1)edge[swap](q3);
\draw[->](q1)edge[swap](q4); \draw[->](q2)edge[swap](q5);
\draw[->](q2)edge[swap](q6); \draw[->](q2)edge[swap](q7);
\draw[->](q4)edge[swap,dotted](q8);
\draw[->](q5)edge[swap,dotted](q9);
\draw[->](q7)edge[swap,dotted](q10);

\node at (-.85,-.6) {$1$};
\node at (.2,-.6) {$i$};
\node at (.85,-.6) {$\ell$};
\node at (-2.4,-2.2) {$1$};
\node at (-1.3,-2.2) {$i$};
\node at (-.55,-2.2) {$\ell$};
\end{tikzpicture}
\vspace*{-0.7cm}
\caption{\small The labeled execution tree of \cref{equ:mpp} on input $\xx$}\label{fig:exe-tree}
\end{figure} }

Now we consider paths in the execution tree. Let $\Sigma=\{1,2,\ldots,l\}$ denote the set of indices of the assignments $\bA_{i}$ ($1\leq i\leq l$). For any finite string $\sigma=a_1 a_2\cdots a_s\in\Sigma^*$, we write $\abs{\sigma}=s$ for its length and $\Pre(\sigma)=\{a_1\cdots a_i\mid i=0,1,\ldots,s-1\}$ for the set of its proper prefixes. The concatenation of two strings $\sigma$ and $\tau=b_1\cdots b_t$ is written as $\sigma\tau=a_1\cdots a_sb_1\cdots b_t$. For simplicity, we let $\bA_\sigma=\bA_{a_s}\circ\cdots\circ \bA_{a_2}\circ \bA_{a_1}$ and $\bA_\epsilon=\mathbf{id}$ (i.e $\bA_\epsilon(\bx)\equiv\bx$) for the empty string $\epsilon$. Then $\bA_{\sigma\tau}=\bA_{\tau}\circ\bA_{\sigma}$. Similarly, for an infinite sequence $\sigma=a_1 a_2\cdots\in\Sigma^\omega$, we define $\abs{\sigma}=\infty$ and $\Pre(\sigma)=\{a_1\cdots a_i\mid i=0,1,\ldots\}$. The concatenation of a finite string $\tau=b_1\cdots b_t\in\Sigma^*$ and an infinite string $\sigma\in\Sigma^\omega$ is defined as $\tau\sigma=b_1\cdots b_ta_1a_2\cdots\in\Sigma^\omega$.

Then any finite or infinite path from the root in the execution tree can be identified by a finite or infinite string over $\Sigma$. Specifically, for any $\sigma=a_1a_2\cdots\in\Sigma^*\cup\Sigma^\omega$, the corresponding execution path is
\[\bx\xrightarrow{\bA_{a_1}}\bA_{a_1}(\bx)\xrightarrow{\bA_{a_2}}\bA_{a_2}(\bA_{a_1}(\bx))\xrightarrow{\bA_{a_3}}\cdots.\]%
Moreover, any node in the execution tree is of the form $\bA_\sigma(\bx)$, where $\sigma\in\Sigma^*$ represents the execution history. Its ancestor nodes are $\bA_\tau(\bx)$ (with $\tau\in\Pre(\sigma)$). According to the definition of the execution tree, we have $G(\bA_\tau(\bx))=0$. Then all paths in the execution tree are given as follows.
\begin{defn}[Execution Paths \cite{LXZZ14}]
The set of \emph{execution paths} of \cref{equ:mpp} for an input $\bx\in \mathbb K^d$ is defined as
\[\Path(\bx) \triangleq \{\sigma\in\Sigma^*\cup\Sigma^\omega\mid \forall \tau\in \Pre(\sigma)\colon G(\bA_\tau(\bx))=0\}.\]
\end{defn}
For any path $\sigma$, we write the set of corresponding polynomials as $T_\sigma^- \triangleq \{G\circ \bA_\tau\mid \tau\in \Pre(\sigma)\}$. Then it is obvious that
\[
\sigma\in\Path(\bx)\iff \bx\in Z(T_\sigma^-).
\]

\subsection{Termination of MPPs}
Now we formalize the meaning of ``termination'' of \cref{equ:mpp}. Intuitively, ``a program terminates'' means that its execution will be accomplished within a finite number of steps, indicating that the execution tree is finite (namely, with only a finite number of nodes). Formally, we have
\begin{defn}[Termination]\label{def:Termi}\noindent
\begin{enumerate}\item For an input $\bx\in \bbK^d$, \cref{equ:mpp} is \emph{terminating} if $\abs{\Path(\bx)}<\infty$, i.e., $\Path(\bx)$ is a finite set; it is \emph{non-terminating} otherwise.
\item The \emph{set of non-terminating inputs} (\textbf{NTI}) of \cref{equ:mpp} is defined as $D \triangleq \{\bx\in\bbK^d\mid \abs{\Path(\bx)}=\infty\}.$
\end{enumerate}\end{defn}

By K\"{o}nig's lemma \cite{koenig}, the execution tree is infinite if and only if it contains an infinite path, i.e.,
\[
\abs{\Path(\bx)}=\infty \iff \Path(\bx)\cap\Sigma^\omega\neq\emptyset
\]
for all $\bx\in\bbK^d$. The \textbf{NTI} can thus be expressed as  
\begin{equation}\label{equ:D}D=\bigcup\nolimits_{\sigma\in\Sigma^\omega}Z(T_\sigma^-).\end{equation}
\oomit{
in another way.
\begin{prop}
\end{prop}
\begin{proof} It directly follows from \cref{equ:path-Z,equ:konig}. \end{proof} }

We are further interested in program termination within a fixed number of iterations:
\begin{defn}[$n$-Termination]\label{def:n-Termi}\noindent
\begin{enumerate}\item \Cref{equ:mpp} is \emph{$n$-terminating} for an input $\bx\in \bbK^d$, if $\abs{\sigma}\leq n$ for all $\sigma\in\Path(\bx)$.
\item The \emph{set of $n$-non-terminating inputs} ($n$-\textbf{NTI}) of \cref{equ:mpp} is defined as
$D_n \triangleq \{\bx\in\bbK^d\mid \exists \sigma\in\Path(\bx)\colon \abs{\sigma}>n\}.$
\end{enumerate}\end{defn}

The $n$-\textbf{NTI} can be expressed in a similar way to \cref{equ:D}. For clarity, we set $T_\sigma \triangleq T_\sigma^-\cup\{G\circ\bA_\sigma\}$, then it is easy to verify that $T_{\sigma a}^-=T_{\sigma}$ for any $\sigma\in\Sigma^*$ and $a\in\Sigma$. Obviously, we have
\begin{prop}
	\begin{equation}\label{equ:Dn}
	D_n=\bigcup\nolimits_{\abs{\sigma}=n}Z(T_\sigma).
	\end{equation}
\end{prop}
\oomit{\begin{proof}It suffices to note that
\begin{equation*}\begin{split}&\exists\sigma\in\Path(\bx):\ |\sigma|>n\\
\iff&\exists\sigma\in\Path(\bx)\colon |\sigma|=n+1\\
\iff&\exists a\in\Sigma,\sigma\in\Sigma^n\colon \bx\in Z(T_{\sigma a}^-)=Z(T_\sigma).
\end{split}\end{equation*}
Here, the first equivalence is from the fact that $\sigma\in \Path(\bx)\implies \Pre(\sigma)\subseteq \Path(\bx)$; and the second one is from \cref{equ:path-Z}. \end{proof} }

%% file: decidability.tex
\section{Decidability of the termination problem}\label{sec:Deci}
We prove that it is decidable whether an MPP of the form \labelcref{equ:mpp} is terminating on a given input $\bx$. This is done by an algorithm computing the NTI $D$ for an MPP, and thus it suffices to decide whether $\bx\in D$.
\subsection{Characterization of the NTI}
We first investigate the mathematical structure of the NTI, which will be shown to imply the decidability of
the termination problem. 


\begin{defn}[Backward Subset]
For any $X\subseteq \bbK^d$, its \emph{backward subset} under \cref{equ:mpp} is
$\Back(X)\triangleq\bigcup_{a\in\Sigma}(X\cap{\bA_a}^{-1}(X)),$
i.e., $\{x\in X \mid \exists a\in \Sigma\colon \bA_a(x)\in X\}$.
\end{defn}

\begin{lem} \label{lem:2}
For any $n\geq 0$, $\Back(D_n)=D_{n+1}$.
\end{lem}

\begin{proof}
	First, we prove that $D_{n+1}\subseteq \Back(D_n)$. For any $\bx\in D_{n+1}$, it follows from \cref{equ:Dn} that $\bx\in Z(T_\sigma)$ for some $\sigma=a_1\cdots a_{n+1}\in\Sigma^{n+1}$. Then by definition $\bA_{a_1}(\bx)\in Z(T_{a_2\cdots a_{n+1}})\subseteq D_n$. Note that $\bx\in D_{n+1}\subseteq D_n$, we have $\bx\in D_n\cap\bA_{a_1}^{-1}(D_n)\subseteq \Back(D_n)$.
	
	Conversely, to show that $\Back(D_n)\subseteq D_{n+1}$, suppose $\bx\in \Back(D_n)$. We have $\bx\in D_n$ and $\bx\in\bA_a^{-1}(D_n)$ for some $a\in\Sigma$. Then $G(\bx)=0$ and $\bA_a(\bx)\in Z(T_\sigma)$ for some $\sigma\in\Sigma^n$. As $\{G\}\cup\{f\circ \bA_a\mid f\in T_\sigma\}=T_{a\sigma}$, we obtain $\bx\in Z(T_{a\sigma})\subseteq D_{n+1}$.
	This completes the proof.
\end{proof}

A direct consequence of \cref{lem:2} is that the descending chain is strict, namely,
\begin{cor}\label{cor:strict}
$D_0\supset D_1\supset\cdots\supset D_N=D_{N+1}=\cdots$.
\end{cor}

We denote in the sequel by $L(d,a,b)$ the obtained maximal length $L(d,f)$ of polynomial ascending chains under a specific control function $f(i)=ab^{i-1}$, which will play a key role in representing the complexity of several algorithms in the forthcoming discussion.

\begin{thm}\label{thm:Dn_D}\noindent
\begin{enumerate}
\item\label[clause]{ite:algebraic} For any $n\geq 0$, $D_n$ is an algebraic set.
\item\label[clause]{ite:chain} The algebraic sets $D_i$ form a descending chain
\[
D_0\supseteq D_1\supseteq\cdots\supseteq D_n\supseteq\cdots,
\]
and there exists $N$ such that $0\leq N \leq L(d,a,b)$ and
 $\forall n\in \mathbb{N}\colon D_N=D_{N+n}$, where $a = \deg(G)$ and $b=\max\{ \deg(\bA_1),\ldots,\deg(\bA_l)\}$.
\item\label[clause]{ite:D=DN} $D=D_N$, i.e., the fixed-point of the chain is exactly the NTI.
\end{enumerate}
\end{thm}

\begin{proof}
	\Cref{ite:algebraic} is directly from \cref{equ:Dn}, and \cref{ite:chain} follows directly from \cref{prop:monomial,cor:strict,thm:bound}.
	To prove \cref{ite:D=DN}, it suffices to show $D_N\subseteq D$, since $D\subseteq D_n$ ($\forall n\geq 0$) can be verified from the definition. For any $n\geq N$ and any $\bx\in D_N=D_n$, we have $\exists \sigma\in\Path(\bx)\colon \abs{\sigma}>n$. Therefore, $\abs{\Path(\bx)}=\infty$ and $\bx\in D$.
\end{proof}

\begin{rem}
Computing the bound $N$ is unnecessary to obtain the set $D$ of non-terminating
  inputs, as it can be done by repeatedly testing whether a Gr\"{o}bner basis for $I(D_i)$ is also one for
  $I(D_{i+1})$ with $i=0,1,\ldots$,
  in case the chain is strictly ascending. In general, however, for an ascending chain $I_1, I_2, \cdots$ of ideals of a Noetherian ring, $I_n=I_{n+1}$ does not imply that $I_n$ is a fixed-point of the chain. Fortunately, \cref{cor:strict} guarantees that the ascending chain $I(D_0),I(D_1),\ldots$ (dually, $D_0, D_1,\ldots$) is strict.
  This idea will be exploited in \cref{alg:NTI}.
\end{rem}

\oomit{\begin{proof} By \cref{thm:Dn_D}, it follows $D_0\supset D_1\supset\cdots\supset D_N=D_{N+1}=\Back(D_N)$. So $D_N$ is a fixed point of the function $\Back$ and thus $D_N=\Back^k(D_N)=D_{n+k}$ for all $k\geq 0$. \end{proof}

It follows from \cref{lem:Dn_D,ite:D=DN,cor:strict} that
\begin{thm}\label{thm:main} $D=D_N$, where $N=\min\{n\geq0\mid D_n=D_{n+1}\}$.
\end{thm} }

The rest of this subsection is dedicated to facts that will be used in \cref{subsec:inequality} to tackle MPPs with inequality guards.

A set $X\subseteq\bbK^d$ is said to be \emph{transitive} under \cref{equ:mpp}, if it can be finitely decomposed as $X=X_1\cup X_2\cup\cdots\cup X_n$ satisfying that for any $i\in\{1,2,\ldots,n\}$, there exists $a\in\Sigma$ and $j\in\{1,2,\ldots,n\}$ such that $\bA_a(X_i)\subseteq X_j$.

\begin{lem}\label{lem:trans}
Let $X\subseteq D_0$ be transitive. Then $X\subseteq D$.
\end{lem}

\begin{proof}
	The transitivity of $X=X_1\cup X_2\cup\cdots\cup X_n$ allows existence of two functions $a \colon \{1,2,\ldots,n\}\rightarrow\Sigma$ and $b \colon \{1,2,\ldots,n\}\rightarrow\{1,2,\ldots,n\}$ such that $\bA_{a(i)}(X_i)\subseteq X_{b(i)}$ for all $i\in\{1,2,\ldots,n\}$. Now for any $\bx\in X$, $\bx\in X_i$ for some $i\in\{1,\ldots,n\}$. Let $b_k=b^k(i)$ and $a_k=a(b^k(i))$ for all $k\geq 0$, we have the path
	\[X_i=X_{b_0}\xrightarrow{\bA_{a_0}} X_{b_1}\xrightarrow{\bA_{a_1}} X_{b_2}\xrightarrow{\bA_{a_2}}\cdots\]
	satisfying $\bA_{a_k}(X_{b_k})\subseteq X_{b_{k+1}}$. Then for $\sigma=a_1a_2\cdots\in\Sigma^\omega$ and any $\tau\in\Pre(\sigma)$, we have $\bA_\tau(X_i)=X_{b_{\abs{\tau}}}\subseteq X\subseteq D_0$. So $\bA_\tau(\bx)\in D_0=Z(\{G\})$, and thus $(G\circ\bA_\tau)(\bx)=0$. Therefore $\sigma\in\Path(\bx)$ and $\bx\in D$.
	
	Moreover, we note that $\{b_0,b_1,\ldots\}\subseteq\{1,2,\ldots,n\}$ has at most $n$ elements, then there exists $j\in\{0,1,\ldots,n-1\}$ and $p\in\{1,\ldots,n\}$ such that $b_j=b_{j+p}$. Thus for all $k\geq 0$ and $t\geq j$, $b_{t+kp}=b^{t+kp}(i)=b^t(i)=b_t$. We have
	\[
	\sigma=(a_1a_2\cdots a_{j-1})(a_ja_{j+1}\cdots a_{j+p-1})^\omega.
	\]
\end{proof}

\begin{thm}\label{thm:greatest-transitive}
$D$ is the largest transitive subset of $D_0$.
\end{thm}

\begin{proof}
	According to \cref{lem:trans}, it suffices to prove that $D$ is transitive. We use the irreducible decomposition $D=Y_1\cup Y_2\cup\cdots\cup Y_n$ to prove transitivity.
	For any $i\in\{1,2,\ldots,n\}$, we have
	\[Y_i\subseteq D=\Back(D)\subseteq\bigcup\nolimits_{a\in\Sigma}{\bA_a}^{-1}(D)=\bigcup\nolimits_{a\in\Sigma}\bigcup\nolimits_{j=1}^n\bA_a^{-1}(Y_j).\]
	Note that $Y_i$ is irreducible and for any $a\in\Sigma$ and any algebraic set $X=Z(T)$, $\bA_a^{-1}(X)=Z(\{f\circ\bA_a \mid f\in T\})$ is also algebraic. Then there exists $a\in\Sigma$ and $j\in\{1,2,\ldots,n\}$ such that $Y_i\subseteq{\bA_a}^{-1}(Y_j)$, i.e., $\bA_a(Y_i)\subseteq Y_j$.
\end{proof}

\begin{cor}\label{cor:regular}For any $\bx\in D$, there is an infinite path in $\Path(\bx)$ with a regular form $\sigma=\sigma_0\sigma_1^\omega$,
where $\sigma_0, \sigma_1\in\Sigma^*$, and $\sigma_1$ is a non-empty string.\end{cor}

\subsection{Algorithms and their complexities}
Now we are ready to formally present algorithms for deciding the termination of \cref{equ:mpp}, i.e.,
  algorithms to compute $D$, or more precisely, a set $B$ of generating polynomials of $D$ with $D=Z(B)$. Thus a program terminates on a given input set $X$, a semi-algebraic set defined by a polynomial formula $\phi(\bx)$, 
is equivalent to the unsatisfiability of $\bigwedge_{f\in B}(f=0) \wedge \phi(\bx)$, which is decidable \`{a} la~\cite{Tarski51}. In fact, for an algebraically closed field, one can decide it by computing the radical ideals.

An algorithm for computing $D$ can be directly obtained from \cref{thm:Dn_D}. Observe that, \cref{equ:Dn} implies that $\prod_{\sigma\in\Sigma^n}T_\sigma$ is a set of generating polynomials of $D_n$ for every $n\in\bbN$. Hence, if we find the number $N=\min\, \{n\geq0\mid D_n=D_{n+1}\}$, which is bounded by $L(d,a,b)$, then the set of generating polynomials for $D_N=D$ is exactly what we need. The detailed procedure is presented in \cref{alg:NTI}, in which $D_n=D_{n+1}$ is checked by means of the following proposition.

\begin{prop}\label{prop:fixedpoint}
	$D_n=D_{n+1}$ iff $f\in I(Z(T_\sigma))$ for all $\sigma\in\Sigma^n$ and all $f\in\prod_{\tau\in\Sigma^{n+1}}T_\tau$.
\end{prop}

\begin{proof}
	Since $D_n$ is always a superset of $D_{n+1}$, by \cref{equ:Dn} we have
	\[D_n=D_{n+1}\iff \forall \sigma\in\Sigma^n\colon Z(T_\sigma)\subseteq \bigcup\nolimits_{\tau\in\Sigma^{n+1}}Z(T_\tau).\]
	Moreover, $\bigcup_{\tau\in\Sigma^{n+1}}Z(T_\tau)=Z(\prod_{\tau\in\Sigma^{n+1}}T_\tau)$. This completes the proof.
\end{proof}

\begin{algorithm}[t]
\caption{Computing NTI from $n$-NTI\label{alg:NTI}}
\begin{algorithmic}[1]
\Require The dimension $d$, the polynomial $G$ and the polynomial vectors $\bA_1,\ldots,\bA_l$.
\Ensure The integer $N = \min\, \{n\geq0\mid D_n=D_{n+1}\}$ and a basis $S_0$ of $D_N$.
\State{\textbf{set of} polynomial sets $S_0\gets \emptyset$} \Comment{for generating polynomials of $D_{n}$}
\State{\textbf{set of} polynomial sets $S_1\gets \{T_\epsilon\}$} \Comment{for generating polynomials of $D_{n+1}$}
\State{\textbf{bool} $b\gets {\rm False}$}
\State{\textbf{integer} $n\gets -1$}
\While{$\neg b$}
\State $S_0\gets S_1$
\State $S_1\gets \emptyset$
\State $n\gets n+1$
\For{$T_\sigma\in S_0$}
\For{$a\gets 1, l$}
\State $T_{\sigma a}\gets T_\sigma\cup \{G\circ\bA_{\sigma a}\}$
\State $S_1\gets S_1\cup \{T_{\sigma a}\}$
\EndFor
\EndFor
\State $b\gets {\rm True}$ \Comment{test if $D_n=D_{n+1}$}
\For{$T_\sigma\in S_0$}
\For{$f\in\prod S_1$}
\State $b\gets b\wedge (f\in I(Z(T_\sigma)))$
\EndFor
\EndFor
\EndWhile
\State \textbf{return} $n$, $S_0$
\end{algorithmic}
\end{algorithm}

\paragraph{Complexity of \cref{alg:NTI}} By \cref{thm:Dn_D}, the \textbf{while} loop terminates after $N$ iterations, with
$N$ bounded by $L(d,a,b)$.
In the $n$-th iteration, there are mainly two computation steps: the first one is to add the polynomial set $T_\tau$ into $S_1$ for each  $\tau\in\Sigma^{n+1}$ and thus will be executed $O(|\Sigma^{n+1}|)=O(l^{n+1})$ times; the second step concerns checking the condition $f\in I(Z(T_\sigma))$ for all $\sigma\in\Sigma^n$ and all $f\in\prod_{\tau\in\Sigma^{n+1}}T_\tau$, leading to $O(l^n(n+2)^{l^{n+1}})$ times of the membership checking. Thus, the total time complexity is $O(l^N(N+2)^{l^{N+1}})$.

In \cref{alg:NTI}, the set of generating polynomials of $D_n$ is directly constructed as $\prod_{\sigma\in\Sigma^n}T_\sigma$ and is in general a huge set. To find a more efficient algorithm, we consider the generating polynomials defined as
 $B_0\triangleq\{G\}$ and $B_{n+1}\triangleq B_n\cup\prod_{a=1}^l(B_n\circ \bA_a)$ for $n \in \mathbb{N}$. Here $B_n\circ\bA_a\triangleq\{f\circ\bA_a\mid f\in B_n\}$. Then,
 
\begin{prop}\label{prop:recursiveB}
	$D_n=Z(B_n)$ for any $n \in \mathbb{N}$.
\end{prop}

\begin{proof}
	It can be verified that $D_n=Z(B_n)\implies\bA_a^{-1}(D_n)=Z(B_n\circ \bA_a)$. Then by applying \cref{lem:2}, the claim can be proved immediately by induction on $n$.
\end{proof}

The advantage of constructing $B_n$ in this recursive way is that we can compute the reduced Gr\"{o}bner basis of $B_n$ during the procedure such that the generating set of $D_n$ can be kept as small as possible. Using this method, a more efficient algorithm for computing the NTI is given in
\cref{alg:NTIgb}, whose correctness is guaranteed by the following result.

\begin{prop}\label{prop:groebner-correctness}
	$\forall n\in \mathbb{N}\colon D_{\hat{N}}=D_{\hat{N}+n}$ for $\hat{N}\triangleq\min\, \{n\mid B_n=B_{n+1}\}$, and
   $\hat{N} \leq L(d,a,b)$, where $L(d,a,b)$ is the same as in \cref{thm:Dn_D}.
\end{prop}

\begin{proof}
	Note that $B_n\subseteq B_{n+1}$ for all $n$. Thus $\hat{N}$ is well-defined due to \cref{thm:asc-chain}, and	bounded by $L(d,a,b)$ according to \cref{prop:monomial,thm:bound}.
\end{proof}

\begin{algorithm}[t]
\caption{Computing NTI using Gr\"{o}bner bases\label{alg:NTIgb}}
\begin{algorithmic}[1]
\Require{The dimension $d$, polynomial $G$ and polynomial vectors $\bA_1,\ldots,\bA_l$.}
\Ensure{The integer $\hat{N} = \min\, \{n\mid B_n=B_{n+1}\}$ and a Gr\"{o}bner basis $B$ of $D_{\hat{N}}$.}
\State \textbf{set of} polynomials $B \gets \{G\}$
\State \textbf{polynomial} $f \gets 0$
\State \textbf{polynomial} $r \gets 0$
\State \textbf{integer} $n \gets -1$
\State \textbf{bool} $c \gets {\rm True}$
\While{$c$}
	\State $c \gets {\rm False}$
	\State $n \gets n+1$
	\For{$f \in \prod_{i=1}^l(B\circ\bA_i)$}
		\State $r \gets \Rem(f,B)$
		\If{$r \neq 0$}
			\State $B \gets \GB(B \cup \{r\})$
			\State $c \gets {\rm True}$
		\EndIf
	\EndFor
\EndWhile
\State \textbf{return} $n$, $B$
\end{algorithmic}
\end{algorithm}

\paragraph{Complexity of \cref{alg:NTIgb}} There are $\hat{N}$ iterations of the \textbf{while} loop during the execution. In the $n$-th iteration, there are at most $\abs{B_{n+1}}$ polynomials to be added into $B$. Note that $\abs{B_n}=O(2^{l^{n-1}})$. Thus, the time complexity is $O(2^{l^{\hat{N}}})$.

\begin{rem}It is clear by definition that, for the same input, the output $\hat{N}$ of \cref{alg:NTIgb} is at least the output $N$ of \cref{alg:NTI}. Moreover, the example below shows that $\hat{N}$ may be different from $N$:
\begin{equation*}
\textsc{while} \quad (x^2+y^2=0)\quad
  \left\{(x,y) \coloneqq (x,x+y);\right\}.
\end{equation*}
With this loop as input, the output is $N=0$ (for $\bbK=\mathbb{R}$) or $1$ (for $\bbK=\mathbb{C}$), yet $\hat{N}=2$.
\oomit{\item The complexity of the algorithms is expressed by $l$ and $N$ (or $\hat{N}$). Note that $N$ (or $\hat{N}$) is not an input, so it should be further elaborated. Denote by $a$ the degree of $G$, and by $b$ the maximal degree of all polynomials in vectors $\bA_1,\ldots,\bA_l$. Define a function $f$ by $f(n)=ab^n$ for all $n\in\bbN$. Then all polynomials in $B_n$ are of degrees not greater than $f(n)$ and thus $\gdeg\langle B_n\rangle\leq f(n)$. From the results of the next section, we can find a computable function $L(n,f)$ of $d$, $a$ and $b$, such that $N\leq\hat{N}\leq L(d,a,b)$.\end{enumerate}}\end{rem}

%% file: extensions.tex
\section{Extensions and applications} \label{sec:ext}
In this section, we extend the obtained results on MPPs to more general polynomial programs, including polynomial guarded commands and MPPs with inequality guards. We further show potential applications to invariant generation.

\subsection{Polynomial guarded commands}
Consider a more general model of polynomial programs named
\emph{polynomial guarded commands} (PGCs), which are typical guarded commands~\cite{dijkstra75}
with all expressions being polynomials and guards being polynomial equalities: 
\begin{defn}[PGCs with Equality Guards]\label{def:pgc}
A \emph{PGC with equality guards} is of the form
\begin{equation}\label[pgc]{equ:pgc}
    \begin{array}{lcl}
  \textsc{do} &  &  G_1(\bx)=0 \longrightarrow \bx\coloneqq \bA_{1}(\bx);\\
     &  \parallel & G_2(\bx)=0 \longrightarrow \bx\coloneqq \bA_{2}(\bx); \\
 &   &  \quad \quad \vdots\\
  &  \parallel & G_{l-1}(\bx)=0 \longrightarrow \bx\coloneqq \bA_{l-1}(\bx); \\
  &  \parallel & G_l(\bx)=0 \longrightarrow \bx\coloneqq \bA_{l}(\bx); \\
  \textsc{od}, & &
   \end{array}
\end{equation}
where
\begin{enumerate}
    \item $\bx\in \bbK^d$ denotes the vector of program variables;
    \item $ G_i\in \bbK[\bx]$ are polynomials and $G_i({\bx})=0$ are guards. Like MPPs, a PGC admits general guards of the form $\bigvee_{i=1}^M\bigwedge_{j=1}^{N_i} G_{ij}(\bx)=0$;
    \item $ \bA_{i}\in \bbK^d[\bx]$ ($ 1\leq i\leq l$) are vectors of polynomials, describing the updates of program variables.
  \end{enumerate}
\end{defn}

\begin{rems}\noindent
\begin{enumerate}
\item Loosely speaking, given an input $\bx$, whenever some guards $G_i$ are satisfied, one of them is nondeterministically chosen and the corresponding assignment is taken. It repeats until none of the guards holds.
\item Evidently, when all $G_i(\bx)$ are identical, the PGC degenerates to an MPP; if $G_i(\bx)=0\wedge G_j(\bx)=0$ has no real solution for any $i\neq j$, 
    the choice among these updates becomes deterministic.
\item Notice that the deterministic choice derived from \cref{equ:pgc} cannot be used to define a general
deterministic choice
\[\textsc{if}~ G(\bx)=0~ \textsc{then} ~\bx \coloneqq \bA_1(\bx)~ \textsc{else} ~ \bx \coloneqq \bA_2(\bx),\]
 as it is equivalent to
\[\textsc{if}~G(\bx){=}0 \longrightarrow \bx {\coloneqq} \bA_1(\bx) \parallel G(\bx){\neq} 0 \longrightarrow \bx {\coloneqq} \bA_2(\bx)~\textsc{fi},\]
 which contains $G(\bx)\neq 0$. In \cite{Muller-Olm}, M{\"{u}}ller{-}Olm and Seidl proved that the Post Correspondence Problem (PCP) can
 be encoded into the extension of PGCs by allowing polynomial inequations in guards, indicating that the termination problem of such an extension is undecidable in general. 
\end{enumerate}
\end{rems}

Given an input $\bx$, the set of paths starting from $\bx$ is
      \[
      \Path(\bx) \triangleq \{\sigma \in \Sigma^*\cup \Sigma^{\omega} \mid \forall \tau\in \Pre(\sigma),i\in \Sigma\colon 
      \tau\cdot i\in \Pre(\sigma) \implies G_i(\bA_{\tau}(\bx))=0\}.
      \]
      For a path $\sigma \in \Path(\bx)$, the set of corresponding polynomials $T^-_{\sigma}$ is $\{G_i(\bA_{\tau}) \mid \tau\cdot i \in \Pre(\sigma)\}$. Similarly, we have
      \[\sigma \in \Path(\bx) \iff \bx \in Z(T^-_{\sigma}).\]

A PGC is \emph{non-terminating} on a given input $\bx$ if there exists $\sigma \in \Path(\bx)$ s.t. $\abs{\sigma}=\infty$, otherwise it is \emph{terminating}. We denote the set of non-terminating inputs of the PGC by $D$.

Analogous to MPPs, \cref{equ:pgc} is called $n$-terminating on an input $\bx\in \bbK^d$, if $\abs{\sigma}\leq n$ for all $\sigma\in\Path(\bx)$, and the set of $n$-non-terminating inputs of \cref{equ:pgc} is defined as
\[
D_n \triangleq \{\bx\in\bbK^d\mid \exists \sigma\in\Path(\bx)\colon \abs{\sigma}>n\}.
\]

The NTI $D$ and $n$-NTI $D_n$ of a PGC have the properties:
\begin{thm}\label{thm:chain-pgc}\noindent
\begin{enumerate}
\item\label{pgc-ite:algebraic} For any $n\geq 0$, $D_n$ is an algebraic set.
\item\label{pgc-ite:chain} $D_i$, $i\in \mathbb{N}$, form a descending chain of algebraic sets, i.e.,
\[
D_0\supseteq D_1\supseteq\cdots\supseteq D_n\supseteq\cdots,
\]
and there exists a least $M$ such that
\[0\leq M \leq L(d,g,a) \wedge \forall n\in \mathbb{N}\colon D_M=D_{M+n},\]
  where $g$ and $a$ are the maximum degree of $G_i$ and $\bA_i$ respectively, for $ 1\leq i\leq l$.
\item\label{pgc-ite:D=DN} $D=D_M$ and
\[
D_0\supset D_1\supset\cdots\supset D_M=D_{M+1} =\cdots.
\]
\end{enumerate}
\end{thm}

\begin{proof}
	Analogous to that of \cref{thm:Dn_D}.
\end{proof}

Based on \cref{thm:chain-pgc}, algorithms akin to \cref{alg:NTI,alg:NTIgb} for computing the NTI of a given PGC can be obtained without any substantial change.

\subsection{MPPs with inequality guards}\label{subsec:inequality}
The method proposed in \cref{sec:Deci} becomes inapplicable as soon as inequalities are involved in the guards of MPPs. In fact, the termination problem in this case is generally undecidable (even not semi-decidable~\cite{bradley:polynomial}) and thus cannot be completely solved. However, an incomplete method for proving nontermination can be achieved by our results. The basic idea is to strengthen an inequality guard by an equality and then try to prove that the obtained MPP is non-terminating under the strengthened guard.

Consider the following example for the intuition behind.
\begin{equation*}
\verb|ineq| \quad \ \;\textsc{while} \quad (x \ge 0) \quad
  \left\{
   \begin{aligned}
   (x,y) &\coloneqq (x-3,y-3);  \\
   \parallel\, (x,y) &\coloneqq (x-3y,y^2+x^2);  \\
   \parallel\,  (x,y) &\coloneqq (1,0);  \\
    \end{aligned}
   \right\}
\end{equation*}
Denote by \verb|ineq'| the program obtained by replacing $x\ge 0$ with $x-y^2=0$ in \verb|ineq|. Obviously,  $x-y^2=0 \implies x \ge 0$, and thus \verb|ineq| is non-terminating for every non-terminating input of \verb|ineq'|. By our method one can easily verify that with input $(x_0,y_0)=(1,-1)$ and $(4,2)$, \verb|ineq| does not terminate.

The key of proving nontermination using this method is, for a given input $\bx$, finding appropriate strengthened polynomial equalities. According to \cref{cor:regular}, the set of all possible polynomials $G$ can be written as $I=\bigcup_{\sigma_0,\sigma_1\in\Sigma*}I_{\sigma_0,\sigma_1}$, where $I_{\sigma_0,\sigma_1}\,{\triangleq}\,\{G\in\bbK[\bx]\mid \sigma_0\sigma_1^\omega\in\mathrm{Path}(\bx)\}$. Then it suffices to enumerate the polynomials in $I$, until one stronger than the original guard is found or the execution times out. Note that for any $\sigma_0$ and $\sigma_1$, $I_{\sigma_0,\sigma_1}$ can be computed by the invariant-generation procedure proposed below.

\subsection{Invariant generation for MPPs}
Now consider a variant problem: Given vectors of polynomials $\bA_i$ and an algebraic set $D$, how to discover all \emph{polynomial invariants} of the form $G(\bx)=0$ for the MPP with $\textit{true}$ as the loop guard and $\bx \coloneqq \bA_1 \parallel \cdots \parallel \bx \coloneqq \bA_l$ as the loop body w.r.t.~the input set $D$? When $l=1$, it is actually the dual problem to termination: A subset $D$ of the NTI is given, while the polynomial $G$ in the loop guard is unknown and needs to be found.

As argued in \cite{rodriguez:simple}, generating polynomial equality invariants for MPPs is challenging and of interest, with promising applications in the verification of polynomial programs. In the literature, however, only a partial solution to the problem is available, which requires the so-called \emph{solvability assumption}~\cite{rodriguez:simple}. In solvable MPPs, assignments $\bA_i$ are essentially linear. In what follows, we show that by the approach proposed in this paper one can solve this problem
without the solvability assumption (provided a bound on the degree of $G(\bx)$ instead). That is, given an MPP and any integer $k$, we can generate all polynomial invariants of
the form $G(\bx)=0$ with $\deg(G)\leq k$ for the MPP.

Consider an MPP with polynomial assignments $\bA_i$, with $1\leq i\leq l$. Suppose $a=\max\{\deg(\bA_1), \ldots, \deg(\bA_l)\}$. For any integer $k$, let $\Sigma_{k}=\{\sigma \mid \abs{\sigma}\leq L(d,k,a)\}$.
\begin{defn}[$k$-Invariants]
The set of \emph{$k$-invariants} of the MPP is defined as
$\Inv_k(D) \triangleq \{G\in\bbK[\bx]\mid\deg(G)\leq k,\ G(\bA_\sigma(\bx))=0\ \text{for all}\ \bx\in D,\ \sigma\in \Sigma_k\}$.
\end{defn}

By \cref{thm:Dn_D}, the following result is immediate.
\begin{prop}
For any $G\in \Inv_k(D)$, the MPP with $G(\bx)=0$ as the loop guard and
  $\bA_i$ as assignments does not
terminate for any input in $D$.
\end{prop}

Note that $\Inv_k(D)$ is a linear space over $\bbK$, and one can readily find an algorithm to compute a basis of it using constraint-solving from the definition. The following theorem shows that the invariant generation of the MPP can be achieved by enumerating all $\Inv_k(D)$ over $k$.
\begin{thm}The set of all polynomial equality invariants of MPPs with input set $D$ is exactly $\bigcup_k \Inv_k(D)$.\end{thm}

%% file: experiments.tex
\section{Implementation and experiments}\label{sec:case}

We have implemented the proposed algorithms
in \textsc{Maple}\footnote{The codes and experimental results as available at \url{https://github.com/Chenms404/MPPs}.} for discovering non-terminating inputs for MPPs. The implementation takes an MPP as input, and gives the minimum integer $N$ such that $D_N$ is a fixed-point of the descending chain of algebraic sets. In particular, essential computations of Gr\"{o}bner bases and membership of ideals attribute to the package \verb|Groebner| and \verb|PolynomialIdeals|, respectively. In what follows, we demonstrate our approach with examples from the literature, which have been evaluated on a 2.4GHz Intel Core-i5 processor with 16GB RAM running macOS Catalina.

\begin{rem}
Notice that all the experiments are conducted mainly for the purpose of testing and demonstrating the proposed algorithms. A fully automated tool with applications to real programs is not provided, due to the high complexity of the underlying procedures: the membership problem of polynomial ideals is EXPSPACE-complete in the size of the problem instance~\cite{idealmembership89}, while the computation of Gr\"{o}bner bases has even higher worst-case complexity. The same issue applies already to the incomplete method~\cite{LXZZ14} that only approximates the NTI and checks a group of sufficient (yet not necessary) conditions for termination and/or nontermination.
\end{rem}
\vspace*{.2cm}

\noindent \verb|liu2|$\quad$
$\ \,\textsc{while} \quad (x+y=0)\quad
  \left\{
   \begin{aligned}
   (x,y) &\coloneqq (x+1,y-1);  \\
   \parallel\, (x,y) &\coloneqq (x^2,y^2);  \\
    \end{aligned}
   \right\}$ \\[3mm]
\noindent \verb|loop|$\quad$
$\ \,\textsc{while} \quad ((x-2)^2+y^2=0)\quad
  \left\{
   \begin{aligned}
   (x,y) &\coloneqq (x+4,y);  \\
   \parallel\, (x,y) &\coloneqq (x+2,y+1);  \\
    \end{aligned}
   \right\}
 $ \\[3mm]
\noindent \verb|liu3|$\quad$
$\ \,\textsc{while} \quad ((x-1)(x-2)=0)\quad
  \left\{
   \begin{aligned}
   x &\coloneqq 1-x^2;  \\
   \parallel\, x &\coloneqq x+1;  \\
    \end{aligned}
   \right\}$ \\[3mm]
\noindent \verb|prod|$\quad$
$\ \,\textsc{while} \quad ((x-1)^2+(y-1)^2+z^2=0) \quad
  \left\{
   \begin{aligned}
   (x,y,z) &\coloneqq (2x,(y-1)/2,x+z);  \\
   \parallel\, (x,y,z) &\coloneqq (2x,y/2,z);  \\
    \end{aligned}
   \right\}
$ \\[3mm]
\noindent \verb|liu4|$\quad$
$\begin{array}{ll}
\textsc{while} \quad (x-z^3=0 \vee y-z^2=0) \\
  \left\{
   \begin{aligned}
   (x,y,z) &\coloneqq (x,y+2z+1,z+1);  \\
   \parallel\, (x,y,z) &\coloneqq (x-3y+3z-1,y+2z-1,z-1);  \\
    \end{aligned}
   \right\}
\end{array} $ \\[3mm]
\noindent \verb|var4|$\quad$
$\begin{array}{ll}
\textsc{while} \quad (w+x+y+z=0) \\
   \left\{
   \begin{aligned}
   (w,x,y,z) &\coloneqq (w-2,x,y+2z+1,z+w);  \\
   \parallel\, (w,x,y,z) &\coloneqq (w^2,x-3y+3z-1,y+2z-1,z-1);  \\
    \end{aligned}
   \right\}
\end{array}$\\[3mm]
\noindent \Verb!qtwk!$\quad$
$\begin{array}{ll}
\textsc{while} \quad (x_2 = 0) \\
\left\{
\begin{aligned}
\bx &\coloneqq \frac{1}{\sqrt{3}}
\begin{bmatrix}
1 & 1 & 0 & -1\\
1 & -1 & 1 & 0\\
0 &1 & 1 & 1\\
1 & 0 & -1 & 1
\end{bmatrix}
\bx
;  \\
\parallel\, \bx &\coloneqq \frac{1}{\sqrt{3}}
\begin{bmatrix}
1 & 1 & 0 & 1\\
-1 & 1 & -1 & 0\\
0 & 1 & 1 & -1\\
1 & 0 & -1 & -1
\end{bmatrix}
\bx
;  \\
\end{aligned}
\right\} \quad \text{with} \quad \bx = [x_0, x_1, x_2, x_3]^\trans
\end{array}$\vspace*{4.5mm}

\begin{table}[t]
\caption{Evaluation results}\label{tab1}
\vspace*{-.1in}
\begin{center}
\begin{tabular}{llcccrr}
  \toprule
  \multirow{2}{*}{\begin{minipage}{1cm}
      Name
    \end{minipage}}
  & \multirow{2}{*}{\begin{minipage}{1cm}
      Source
    \end{minipage}}
  & \multirow{2}{*}{\begin{minipage}{0.46cm}\centering
      {\small $d$}
    \end{minipage}}
  & \multirow{2}{*}{\begin{minipage}{0.46cm}\centering
      {\small $l$}
    \end{minipage}}
  & \multirow{2}{*}{\begin{minipage}{0.46cm}\centering
      {\small $N$}
    \end{minipage}}
    & \multicolumn{2}{c}{Time (sec)}\\
  \cmidrule(lr){6-7}
  & & & & & \cref{alg:NTI} & \cref{alg:NTIgb}
  \\ \hline
  \verb|overview| & [$\,\star\,$] & 2 & 2 & 1 & 0.244 & 0.012 \\
  \verb|liu1| & \cite{LXZZ14} & 2 & 2 & 0 & 0.006 & 0.002 \\
  \verb|liu2| & \cite{LXZZ14} & 2 & 2 & 0 & 0.003 & 0.002 \\
  \verb|loop| & \cite{CousotCousot77-1} & 2 & 2 & 2 & 90.881 & 0.013 \\
  \verb|liu3| & \cite{LXZZ14} & 1 & 2 & 2 & 121.418 & 0.009 \\
  \verb|ineq'| & [$\,\star\,$] & 2 & 3 & 2 & TO & 0.056 \\
  \verb|prod| & \cite{rodriguez:simple} & 3 & 2 & 4 & TO & 0.948 \\
  \verb|liu4| & \cite{LXZZ14} & 3 & 2 & 4 & TO & 0.515 \\
  \verb|var4| & [$\,\star\,$] & 4 & 2 & 5 & TO & 4.014 \\
  \Verb!qtwk! & \cite{LY14} & 4 & 2 & 1 & 0.438 & 0.036 \\
  \bottomrule
\end{tabular}\\
\end{center}
\vspace*{.1cm}
\begin{minipage}{\columnwidth}
	{\small{Legends: the first two columns specify the name of the program, and the citation from where the program was adapted ([$\,\star\,$] marks examples developed by the authors). $d$ indicates the number of variables in the loop, $l$ represents the number of branches in the loop body, and the last two columns give the time (in seconds) taken by the Algorithms. Timeouts (TO) here are set to 20 minutes. In addition, $\hat{N}=N$ holds for all the examples.}}
\end{minipage}
\vspace*{-0.15in}
\end{table}

The following observations can be drawn from the experimental results collected in \cref{tab1}:
\begin{enumerate}
\item The implementation of \cref{alg:NTIgb} in \textsc{Maple} works efficiently: it outperforms \cref{alg:NTI} and solves all the examples (albeit small) successfully within 6 seconds in total. \Cref{alg:NTI}, however, tends to be strenuous when $l$ or the expected $N$ reaches 2, and gets timeouts for cases with $N \ge 3$.
\item $N$ is roughly equal to $d$ in these examples, at least not getting too large in terms of $d$, which well explains that despite with a worst-case complexity doubly exponential in $N$, \cref{alg:NTIgb} may exhibit a promising performance in practice.
\item Example \Verb!qtwk! is adapted from a nondeterministic quantum program in~\cite{LY14} that models a quantum walk on an undirected graph of four vertices. Our approach succeeds in computing the NTI, which coincides with that in~\cite{LY14}, and thus provides a proof of nontermination.
\oomit{\item Note in Example \verb|ineq| that $x-y^2=0 \implies x \ge 0$, thus the generated NTI are also the non-terminating inputs of the program where we replace the condition $x-y^2=0$ by $x \ge 0$. This demonstrates that even for more general loops with inequality conditions, our approach provides an effective way to find the terminating counterexamples.}
\end{enumerate}

It is also worth highlighting that, as explained in~\cite{LXZZ14}, Example \Verb!liu2! and \Verb!liu3! cannot be handled by the technique thereof based on under/over-approximating the NTI, since no information about termination can be inferred from the approximations, whereas they are successfully solved by the approach developed in this paper. The same argument applies to other examples like \Verb!overview!. A detailed comparison with the approach in~\cite{LXZZ14} in terms of applicability and performance is unfortunately not impossible as no publicly available implementation is provided therein.

%% file: conclusion.tex
\section{Conclusions}\label{sec:conc}

We showed that the termination problem of a family of multi-path polynomial programs with equality conditions is decidable. This class of (multi-path) polynomial programs, to the best of our knowledge, is hitherto the largest fragment with a decidable termination problem: any extension of it by allowing inequalities or inequations in the guards may render the termination problem undecidable. Two variant algorithms for computing the set of non-terminating inputs, and thus deciding termination, are proposed, whose effectiveness was demonstrated on a collection of typical examples from the literature. To analyze complexities of the decision algorithms, we proposed a constructive method to compute the maximal length of Hilbert ascending chains of polynomial ideals under a control function. An explicit recursive function is then obtained for computation, which further implies the Ackermannian complexity of the length, thereby answering the questions raised by Seidenberg. We discussed potential extensions to more general polynomial programs including PGCs and MPPs with inequality guards. A relatively complete approach (in terms of a bound on the degree of polynomial equalities) to synthesize polynomial equality invariants for MPPs was also presented.

For future work, it will be of particular interest to consider invariant generation without a priori specified degrees. Note that even for the simplest case, namely, with a single variable $x$ (i.e., $d=1$) and a unique assignment $A(x)$ (i.e., $l=1$) in the MPP, the problem is non-trivial. In this case, a polynomial equality invariant $G(x)=0$ exists if and only if the input state $x_0$ is a pre-periodical point of $A$. We note that the problem of deciding whether a rational point is pre-periodical for a given polynomial map is related to the Uniform Boundedness Conjecture~\cite{MS94}.